\RequirePackage{fix-cm}
\documentclass[smallextended]{svjour3}       
\smartqed  
\usepackage{graphicx}
\usepackage{amsfonts,bbm,bm,braket}
\usepackage{scalerel}
\usepackage{enumerate}
\usepackage{mathtools}
\usepackage{ifthen}
\usepackage{tensor}
\usepackage{tikz}
\usepackage{tikz-network}
\usetikzlibrary{patterns,decorations.pathreplacing}
\usepackage[colorlinks,bookmarks=false,citecolor=NavyBlue,linkcolor=OliveGreen,url color=blue]{hyperref}

\usepackage{color}
\definecolor{myred}{RGB}{232,102,102}
\definecolor{myblue}{RGB}{187,187,255}
\definecolor{myorange}{RGB}{202,52,51}
\definecolor{mygrey}{RGB}{105,105,105}
\definecolor{OliveGreen}{RGB}{85,107,47}
\definecolor{NavyBlue}{RGB}{0,0,128}
\definecolor{mygreen}{RGB}{34,139,34}
\definecolor{myY}{RGB}{220,255,203}
\definecolor{myYO}{RGB}{255, 220, 151}
\definecolor{myblue1}{RGB}{176,223,229}
\definecolor{myblue2}{RGB}{0,0,128}
\definecolor{myblue3}{RGB}{0,108,255}
\definecolor{myblue4}{RGB}{101,147,245}
\definecolor{myblue5}{RGB}{115,194,251}
\definecolor{myblue6}{RGB}{87,160,211}
\definecolor{myblue7}{RGB}{137,207,240}
\definecolor{myblue8}{RGB}{29,41,81}
\definecolor{myblue9}{RGB}{14,77,146}
\definecolor{myblue10}{RGB}{15,82,186}
\definecolor{myred1}{RGB}{255,36,0}
\definecolor{myred2}{RGB}{205,92,92}
\definecolor{myred3}{RGB}{178,34,34}
\definecolor{myred4}{RGB}{164,90,82}
\definecolor{myred5}{RGB}{255,8,0}
\definecolor{myred6}{RGB}{202,52,51}
\definecolor{myred7}{RGB}{66,13,9}
\definecolor{myred8}{RGB}{141,2,31}
\definecolor{myred9}{RGB}{250,128,114}
\definecolor{myred10}{RGB}{237,41,57}
\definecolor{myyellow1}{RGB}{254,220,86}
\definecolor{myyellow2}{RGB}{255,229,180}
\definecolor{myyellow3}{RGB}{238,220,130}
\definecolor{myyellow4}{RGB}{253,165,15}
\definecolor{myyellow5}{RGB}{255,195,11}
\definecolor{myyellow6}{RGB}{218,165,32}
\definecolor{myyellow7}{RGB}{255,211,0}
\definecolor{myyellow8}{RGB}{248,222,126}
\definecolor{myyellow9}{RGB}{245,245,220}
\definecolor{myyellow10}{RGB}{248,228,115}

\definecolor{mygray1}{RGB}{246,246,246}
\definecolor{mygray2}{RGB}{32,32,32}
\definecolor{mygray3}{RGB}{64,64,64}
\definecolor{mygray4}{RGB}{96,96,96}
\definecolor{mygray5}{RGB}{128,128,128}
\definecolor{mygray6}{RGB}{160,160,160}
\definecolor{mygray7}{RGB}{224,224,224}
\definecolor{mygray8}{RGB}{180,180,180}

\usepackage{tensor}
\newcommand{\be}{\begin{equation}}
\newcommand{\ee}{\end{equation}}
\newcommand{\bea}{\begin{eqnarray}}
\newcommand{\eea}{\end{eqnarray}}
\newcommand{\ba}{\begin{aligned}}
\newcommand{\ea}{\end{aligned}}
\newcommand{\bw}{\begin{widetext}}
\newcommand{\ew}{\end{widetext}}
\newcommand{\1}{\mathbbm{1}}

\def\iid{{\em i.i.d. }}

\newcommand{\Wgategreen}[2]{
\draw[very thick] (#1-0.5, #2 +0.5) -- (#1+0.5,#2-0.5);
\draw[very thick] (#1-0.5,#2-0.5) -- (#1+0.5,#2+0.5);
\draw[ thick, fill=mygreen, rounded corners=2pt] (#1-0.25,#2+0.25) rectangle (#1+0.25,#2-0.25);
\draw[thick] (#1,#2+0.15) -- (#1+0.15,#2+0.15) -- (#1+0.15,#2);
}

\newcommand{\Wgateolivegreen}[2]{
\draw[very thick] (#1-0.5, #2 +0.5) -- (#1+0.5,#2-0.5);
\draw[very thick] (#1-0.5,#2-0.5) -- (#1+0.5,#2+0.5);
\draw[ thick, fill=OliveGreen, rounded corners=2pt] (#1-0.25,#2+0.25) rectangle (#1+0.25,#2-0.25);
\draw[thick] (#1,#2+0.15) -- (#1+0.15,#2+0.15) -- (#1+0.15,#2);
}

\begin{document}

\title{Random Matrix Spectral Form Factor of Dual-Unitary Quantum Circuits
}

\authorrunning{Bruno Bertini, Pavel Kos, Toma\v z Prosen} 

\author{B. Bertini \and P. Kos \and T. Prosen}

\institute{Bruno Bertini \at
              Theoretical Physics, Oxford University, Parks Road, Oxford OX1 3PU, UK,
              \email{bruno.bertini@physics.ox.ac.uk}                                       
           \and
            Pavel Kos \at
              Faculty of Mathematics and Physics, University of Ljubljana, Jadranska 19, SI1000 Ljubljana, Slovenia,
              \email{pavel.kos@fmf.uni-lj.si}
            \and
            Toma\v z Prosen\at  
             Faculty of Mathematics and Physics, University of Ljubljana, Jadranska 19, SI1000 Ljubljana, Slovenia,
              \email{tomaz.prosen@fmf.uni-lj.si}}

\date{Received: date / Accepted: date}

\maketitle

\begin{abstract}
We investigate a class of local quantum circuits on chains of $d-$level systems (qudits) that share the so-called `dual unitarity' property. In essence, the latter property implies that these systems generate unitary dynamics not only when propagating in time, but also when propagating in space. We consider space-time homogeneous (Floquet) circuits and perturb them with a quenched single-site disorder, i.e.\ by applying independent single site random unitaries drawn from arbitrary non-singular distribution over ${\rm SU}(d)$, e.g.\ one concentrated around the identity, after each layer of the circuit. We identify the spectral form factor at time $t$ in the limit of long chains as the dimension of the commutant of a finite set of operators on a qudit ring of $t$ sites. For general dual unitary circuits of qubits $(d=2)$ and a family of their extensions to higher $d>2$, we provide an explicit construction of the commutant and prove that spectral form factor exactly matches the prediction of circular unitary ensemble for all $t$, if only the local 2-qubit gates are different from a SWAP (non-interacting gate). We discuss and partly prove possible extensions of our results to weaker (more singular) forms of disorder averaging, as well as to quantum circuits with time-reversal symmetry, and for computing higher moments of the spectral form factor.

\end{abstract}

\section{Introduction: Quantum chaos conjecture and many-body systems}
\label{intro}

The ubiquitousness of {\em random matrix theory} (RMT) \cite{Mehtabook,sarnak,forrester} descriptions for a diverse range of phenomena in Nature and Society is the example par excellence of effectiveness of mathematics. In quantum dynamical systems, the fact that the fluctuations in the  spectra of unitary evolution operators can be described in terms of structureless ensembles of RMT, characterised solely by unitary and anti-unitary symmetries, has been identified as a defining property of {\em quantum chaos}. The presence of RMT spectral correlations has been related to Hamiltonian chaos of the corresponding limiting classical dynamical system via the so-called {\em quantum chaos conjecture}\footnote{Sometimes referred to also as Bohigas-Giannoni-Schmit conjecture.} (QCC) \cite{casati,berry,bohigas}, while the absence thereof is linked to integrability or regularity of the corresponding classical motion via the Berry-Tabor conjecture \cite{berrytabor,marklof}. A heuristic proof of QCC in terms of
semiclassical periodic orbit theory has been a decades long tour de force \cite{berry85,sieberrichter,mueller}, while a rigorous proof has so far been possible only in a rather restricted setting of completely connected quantum graphs \cite{weidenmueller}.

In systems which lack a small parameter (e.g. an effective Planck's constant), such as extended (many-body) systems of locally interacting quantum spins or fermions, the mechanisms for the validity of QCC have remained obscure despite a plethora of empirical evidence, see e.g. Refs.~\cite{mila,montambaux,hsu,P99,lea}. For such systems, lacking any meaningful limiting classical chaotic dynamics, the agreement of spectral fluctuations with RMT can be considered as the most versatile definition of {\em quantum chaos} and as a robust empirical method for detection of quantum (non)integrability. 

Recently we proposed a rigorous methodology which lead to the first proof of the emergence of RMT spectral 2-point correlation functions in the thermodynamic limit, for a particular locally interacting chain of quantum spins ${1/2}$~\cite{BKP18}. In this paper, we present a generalisation of such methodology and show that it can be extended to a much broader class of systems. In particular, we reformulate our approach in the general language of local quantum circuits --- which are the standard minimal model of quantum many-body (extended) systems with local interactions~\cite{nahum,chalker,chalker2} --- and show that it leads to exact results (in the thermodynamic limit) whenever the ``local gates" (the unitary matrices encoding the nearest-neighbour interactions of local quantum circuits) are \emph{dual-unitary}, i.e. they generate unitary evolution in both time and space. The method described here applies to quantum circuits of qudits (or arbitrary spins) and can easily account for spatially inhomogeneous interactions. Moreover, in contrast to Ref.~\cite{BKP18}, here we focus on the generic case of systems without anti-unitary symmetries (like time-reversal) while only sketch the extension of the results to generic time-reversal invariant case. In summary, in this paper we identify the key mathematical steps for approaching the problem of characterizing quantum ergodicity \cite{zelditch,alicki} and quantum chaos \cite{tolya,chalker,ljubotina} through spectral correlations in extended quantum spin lattice systems. Although our results hold in a specific setting (and in the thermodynamic limit only), they represent the first rigorous proof of the emergence of RMT behaviour in a class of extended quantum spin systems to the best of our knowledge. They pertain to both the case of quenched disorder and the clean limit, and provide the first proof of the QCC in the many-body realm.

The rest of the paper is laid out as follows. In Sec.~\ref{sec:basic} we introduce the basic concepts and provide their definitions. In Sec.~\ref{sec:main} we state and interpret our main results, while in Sec.~\ref{sec:proofs} we elaborate the proofs. In Sec.~\ref{sec:discussion} we discuss some straightforward extensions and generalisations of our results.
While the treatment of spatially inhomogeneous dual-unitary interactions in Sec.~\ref{Sec:inhom} is rigorous, the extensions of the techniques to study fluctuations (\ref{sec:fluctuations}) and singular disorder distributions (\ref{sec:nonisotropic}) are speculative at this point.

\section{Basic concepts}
\label{sec:basic}

\subsection{Floquet quantum circuits}

In this work we consider a class of quantum many-body systems known as \emph{Floquet local quantum circuits}. They consist of a set of $2L$, $L\in\mathbb N$, quantum variables with $d$ internal states (``qudits''), that can be thought of as
arranged on a 1-dimensional periodic lattice $\Lambda_L=\frac{1}{2}\mathbb Z_{2L}$. The Hilbert space of the system is  
\be
\mathcal H_{2L} = (\mathcal H_1)^{\otimes 2 L} = \mathbb C^{\mathcal N},
\ee
where the ``local Hilbert space'' $\mathcal H_1 = \mathbb C^d$ is the Hilbert space of a single qudit and $\mathcal N = d^{2L}$ is the dimension of $\mathcal H_{2L}$.

In these systems the time evolution is discrete and generated by integer powers of the unitary operator 
\be 
\mathbb U_{L} :=\prod_{x \in \mathbb Z_{L}} \eta_{x,L}(U_{x,1}) \prod_{x \in \mathbb Z_{L}+\frac{1}{2}}\!\!\eta_{x,L}(U_{x,\frac{1}{2}}) \,, \label{eq:Floquet}
\ee
conventionally called the ``Floquet operator''. In writing Eq.~\eqref{eq:Floquet} we introduced the following definitions. 
\begin{itemize}
\item[(i)] We indicated by $\eta_{x,n} \!:\, {\rm End}(\mathcal H_2)\,\rightarrow\,{\rm End}(\mathcal H_{2n})$, with $n\in \mathbb N$ and $x\in \Lambda_{n}$, the linear map defined by 
\be
\eta_{x,n}(O):=\Pi_{2n}^{2x-1} (O \otimes\1_{2(n-1)}) \Pi_{2n}^{-(2x-1)}.
\ee
Here $\mathcal H_n=(\mathcal H_1)^{\otimes n}$ denotes the Hilbert space of a periodic qudit chain of $n$ sites, 
while $\Pi_n$ and $\1_n$ designate respectively the periodic shift operator and the identity operator over $\mathcal H_n$. Explicitly we have  
\be
\Pi_n \ket{j_1}\otimes \ket{j_2}\otimes\cdots\ket{j_n} = \ket{j_n} \otimes \ket{j_1}\otimes \cdots\ket{j_{n-1}},
\ee 
with $\Pi^n_n = \1_n$,
where 
\be
\mathcal R = \{\ket{j};\; j=0,\ldots, d-1 \},
\label{eq:realbasis}
\ee
is the canonical orthonormal basis of $\mathcal H_1=\mathbb C^d$.
\item[(ii)] We introduced the function $U_{\cdot, \cdot} : \Lambda_L \times  \frac{1}{2}\mathbb Z_2\,\rightarrow\,{\rm U}(d^2)$, where ${\rm U}(N)$ denotes the group of $N\times N$ unitary matrices. The operators $U_{x,\frac{1}{2}}, U_{x,1}\equiv U_{x,0}\in{\rm U}(d^2)$ define the interaction among neighbouring qudits at sites $x-\frac{1}{2},x$ for half-odd integer and integer times respectively. These operators encode all physical information about the dynamics and will be referred to as the ``local gates''. 
\end{itemize}
This setting can be regarded as the simplest possible modelling for extended quantum many-body systems~\cite{chalker,chalker2}. Indeed, it captures what can be considered the primary and most essential feature of an extended system, i.e., the locality of the interactions. It is precisely this feature that distinguishes an extended system from a single quantum variable with arbitrary many internal states, for example, a single (arbitrary high) spin. Note that even though local quantum circuits describe time-depended dynamics, they can also be thought of as approximations of time-independent local interactions obtained through the use of the Suzuki-Trotter decomposition \cite{suzuki,osborne}. Finally, we stress that this precise setting emerges naturally in the context of quantum simulation, for instance, through the use of the recently developed Google's Sycamore processor~\cite{google}.

The time evolution generated by \eqref{eq:Floquet} admits the following convenient graphical representation 
\begin{align}
&\mathbb U_{L}^t=
\begin{tikzpicture}[baseline=(current  bounding  box.center), scale=0.55]
\foreach \i in {0,1,2}{
\draw[thick, dotted] (-9.5,2*\i-1.7) -- (0.4,2*\i-1.7);
\draw[thick, dotted] (-9.5,2*\i-1.3) -- (0.4,2*\i-1.3);
}
\foreach \i in {3,...,13}{
\draw[gray, dashed] (-12.5+\i,-2.1) -- (-12.5+\i,4.3);
}
\foreach \i in {-1,...,5}{
\draw[gray, dashed] (-9.75,3-\i) -- (.75,3-\i);
}
\foreach \i in{1.5,2.5,3.5}{
\draw[thick] (0.5,2*\i-0.5-3.5) arc (45:-90:0.17);
\draw[thick] (-10+0.5+0,2*\i-0.5-3.5) arc (90:270:0.15);
}
\foreach \i in{0.5,1.5,2.5}
{
\draw[ thick] (0.5,1+2*\i-0.5-3.5) arc (-45:90:0.17);
\draw[ thick] (-10+0.5,1+2*\i-0.5-3.5) arc (270:90:0.15);
}
\foreach \i in {1,2}{
\Text[x=1.25,y=-2+2*\i]{\scriptsize$\i$}
}
\foreach \i in {1,3}{
\Text[x=1.25,y=-2+\i]{\small$\frac{\i}{2}$}
}
\foreach \i in {1,3,5}{
\Text[x=-7.5+\i-2,y=-2.6]{\small$\frac{\i}{2}$}
}
\foreach \i in {1,2,3}{
\Text[x=-7.5+2*\i-2,y=-2.6]{\scriptsize${\i}$
}
}
\foreach \jj[evaluate=\jj as \j using -2*(ceil(\jj/2)-\jj/2)] in {-1,-3,-5}{
\foreach \i in {1}
{
\draw[thick] (.5-2*\i-1*\j,-2-1*\jj) -- (1-2*\i-1*\j,-1.5-\jj);
\draw[thick] (1-2*\i-1*\j,-1.5-1*\jj) -- (1.5-2*\i-1*\j,-2-\jj);
\draw[thick] (.5-2*\i-1*\j,-1-1*\jj) -- (1-2*\i-1*\j,-1.5-\jj);
\draw[thick] (1-2*\i-1*\j,-1.5-1*\jj) -- (1.5-2*\i-1*\j,-1-\jj);
\draw[thick, fill=myred5, rounded corners=2pt] (0.75-2*\i-1*\j,-1.75-\jj) rectangle (1.25-2*\i-1*\j,-1.25-\jj);
\draw[thick] (-2*\i+2,-1.35-\jj) -- (-2*\i+2.15,-1.35-\jj) -- (-2*\i+2.15,-1.5-\jj);%
}
\foreach \i in {2}
{
\draw[thick] (.5-2*\i-1*\j,-2-1*\jj) -- (1-2*\i-1*\j,-1.5-\jj);
\draw[thick] (1-2*\i-1*\j,-1.5-1*\jj) -- (1.5-2*\i-1*\j,-2-\jj);
\draw[thick] (.5-2*\i-1*\j,-1-1*\jj) -- (1-2*\i-1*\j,-1.5-\jj);
\draw[thick] (1-2*\i-1*\j,-1.5-1*\jj) -- (1.5-2*\i-1*\j,-1-\jj);
\draw[thick, fill=myred4, rounded corners=2pt] (0.75-2*\i-1*\j,-1.75-\jj) rectangle (1.25-2*\i-1*\j,-1.25-\jj);
\draw[thick] (-2*\i+2,-1.35-\jj) -- (-2*\i+2.15,-1.35-\jj) -- (-2*\i+2.15,-1.5-\jj);%
}
\foreach \i in {3}
{
\draw[thick] (.5-2*\i-1*\j,-2-1*\jj) -- (1-2*\i-1*\j,-1.5-\jj);
\draw[thick] (1-2*\i-1*\j,-1.5-1*\jj) -- (1.5-2*\i-1*\j,-2-\jj);
\draw[thick] (.5-2*\i-1*\j,-1-1*\jj) -- (1-2*\i-1*\j,-1.5-\jj);
\draw[thick] (1-2*\i-1*\j,-1.5-1*\jj) -- (1.5-2*\i-1*\j,-1-\jj);
\draw[thick, fill=myred3, rounded corners=2pt] (0.75-2*\i-1*\j,-1.75-\jj) rectangle (1.25-2*\i-1*\j,-1.25-\jj);
\draw[thick] (-2*\i+2,-1.35-\jj) -- (-2*\i+2.15,-1.35-\jj) -- (-2*\i+2.15,-1.5-\jj);%
}
\foreach \i in {4}
{
\draw[thick] (.5-2*\i-1*\j,-2-1*\jj) -- (1-2*\i-1*\j,-1.5-\jj);
\draw[thick] (1-2*\i-1*\j,-1.5-1*\jj) -- (1.5-2*\i-1*\j,-2-\jj);
\draw[thick] (.5-2*\i-1*\j,-1-1*\jj) -- (1-2*\i-1*\j,-1.5-\jj);
\draw[thick] (1-2*\i-1*\j,-1.5-1*\jj) -- (1.5-2*\i-1*\j,-1-\jj);
\draw[thick, fill=myred2, rounded corners=2pt] (0.75-2*\i-1*\j,-1.75-\jj) rectangle (1.25-2*\i-1*\j,-1.25-\jj);
\draw[thick] (-2*\i+2,-1.35-\jj) -- (-2*\i+2.15,-1.35-\jj) -- (-2*\i+2.15,-1.5-\jj);%
}
\foreach \i in {5}
{
\draw[thick] (.5-2*\i-1*\j,-2-1*\jj) -- (1-2*\i-1*\j,-1.5-\jj);
\draw[thick] (1-2*\i-1*\j,-1.5-1*\jj) -- (1.5-2*\i-1*\j,-2-\jj);
\draw[thick] (.5-2*\i-1*\j,-1-1*\jj) -- (1-2*\i-1*\j,-1.5-\jj);
\draw[thick] (1-2*\i-1*\j,-1.5-1*\jj) -- (1.5-2*\i-1*\j,-1-\jj);
\draw[thick, fill=myred1, rounded corners=2pt] (0.75-2*\i-1*\j,-1.75-\jj) rectangle (1.25-2*\i-1*\j,-1.25-\jj);
\draw[thick] (-2*\i+2,-1.35-\jj) -- (-2*\i+2.15,-1.35-\jj) -- (-2*\i+2.15,-1.5-\jj);%
}
}
\foreach \jj[evaluate=\jj as \j using -2*(ceil(\jj/2)-\jj/2)] in {-4,-2,0}{
\foreach \i in {1}
{
\draw[thick] (.5-2*\i-1*\j,-2-1*\jj) -- (1-2*\i-1*\j,-1.5-\jj);
\draw[thick] (1-2*\i-1*\j,-1.5-1*\jj) -- (1.5-2*\i-1*\j,-2-\jj);
\draw[thick] (.5-2*\i-1*\j,-1-1*\jj) -- (1-2*\i-1*\j,-1.5-\jj);
\draw[thick] (1-2*\i-1*\j,-1.5-1*\jj) -- (1.5-2*\i-1*\j,-1-\jj);
\draw[thick, fill=myred8, rounded corners=2pt] (0.75-2*\i-1*\j,-1.75-\jj) rectangle (1.25-2*\i-1*\j,-1.25-\jj);
\draw[thick] (-2*\i+1,-1.35-\jj) -- (-2*\i+1.15,-1.35-\jj) -- (-2*\i+1.15,-1.5-\jj);%
}
\foreach \i in {2}
{
\draw[thick] (.5-2*\i-1*\j,-2-1*\jj) -- (1-2*\i-1*\j,-1.5-\jj);
\draw[thick] (1-2*\i-1*\j,-1.5-1*\jj) -- (1.5-2*\i-1*\j,-2-\jj);
\draw[thick] (.5-2*\i-1*\j,-1-1*\jj) -- (1-2*\i-1*\j,-1.5-\jj);
\draw[thick] (1-2*\i-1*\j,-1.5-1*\jj) -- (1.5-2*\i-1*\j,-1-\jj);
\draw[thick, fill=myred10, rounded corners=2pt] (0.75-2*\i-1*\j,-1.75-\jj) rectangle (1.25-2*\i-1*\j,-1.25-\jj);
\draw[thick] (-2*\i+1,-1.35-\jj) -- (-2*\i+1.15,-1.35-\jj) -- (-2*\i+1.15,-1.5-\jj);%
}
\foreach \i in {3}
{
\draw[thick] (.5-2*\i-1*\j,-2-1*\jj) -- (1-2*\i-1*\j,-1.5-\jj);
\draw[thick] (1-2*\i-1*\j,-1.5-1*\jj) -- (1.5-2*\i-1*\j,-2-\jj);
\draw[thick] (.5-2*\i-1*\j,-1-1*\jj) -- (1-2*\i-1*\j,-1.5-\jj);
\draw[thick] (1-2*\i-1*\j,-1.5-1*\jj) -- (1.5-2*\i-1*\j,-1-\jj);
\draw[thick, fill=myred6, rounded corners=2pt] (0.75-2*\i-1*\j,-1.75-\jj) rectangle (1.25-2*\i-1*\j,-1.25-\jj);
\draw[thick] (-2*\i+1,-1.35-\jj) -- (-2*\i+1.15,-1.35-\jj) -- (-2*\i+1.15,-1.5-\jj);%
}
\foreach \i in {4}
{
\draw[thick] (.5-2*\i-1*\j,-2-1*\jj) -- (1-2*\i-1*\j,-1.5-\jj);
\draw[thick] (1-2*\i-1*\j,-1.5-1*\jj) -- (1.5-2*\i-1*\j,-2-\jj);
\draw[thick] (.5-2*\i-1*\j,-1-1*\jj) -- (1-2*\i-1*\j,-1.5-\jj);
\draw[thick] (1-2*\i-1*\j,-1.5-1*\jj) -- (1.5-2*\i-1*\j,-1-\jj);
\draw[thick, fill=myred9, rounded corners=2pt] (0.75-2*\i-1*\j,-1.75-\jj) rectangle (1.25-2*\i-1*\j,-1.25-\jj);
\draw[thick] (-2*\i+1,-1.35-\jj) -- (-2*\i+1.15,-1.35-\jj) -- (-2*\i+1.15,-1.5-\jj);%
}
\foreach \i in {5}
{
\draw[thick] (.5-2*\i-1*\j,-2-1*\jj) -- (1-2*\i-1*\j,-1.5-\jj);
\draw[thick] (1-2*\i-1*\j,-1.5-1*\jj) -- (1.5-2*\i-1*\j,-2-\jj);
\draw[thick] (.5-2*\i-1*\j,-1-1*\jj) -- (1-2*\i-1*\j,-1.5-\jj);
\draw[thick] (1-2*\i-1*\j,-1.5-1*\jj) -- (1.5-2*\i-1*\j,-1-\jj);
\draw[thick, fill=myred1, rounded corners=2pt] (0.75-2*\i-1*\j,-1.75-\jj) rectangle (1.25-2*\i-1*\j,-1.25-\jj);
\draw[thick] (-2*\i+1,-1.35-\jj) -- (-2*\i+1.15,-1.35-\jj) -- (-2*\i+1.15,-1.5-\jj);%
}
}
\Text[x=-2,y=-2.6]{$\cdots$}
\Text[x=0.47,y=-2.6]{\small $L\equiv 0$}
\Text[x=-4,y=-3.35]{\small $x$}
\Text[x=1.25,y=4]{\small $t$}
\Text[x=2,y=1]{\small$\tau$}
\Text[x=1.25,y=3.2]{$\vdots$}
\end{tikzpicture},
\end{align}
where each local gate is represented by
\be
\label{eq:Ugate}
U_{x,\tau}\equiv U_{x,{\rm mod}(\tau,1)}=\begin{tikzpicture}[baseline=(current  bounding  box.center), scale=.7]
\draw[ thick] (-4.25,0.5) -- (-3.25,-0.5);
\draw[ thick] (-4.25,-0.5) -- (-3.25,0.5);
\draw[ thick, fill=myred, rounded corners=2pt] (-4,0.25) rectangle (-3.5,-0.25);
\draw[thick] (-3.75,0.15) -- (-3.75+0.15,0.15) -- (-3.75+0.15,0);
\Text[x=-4.25,y=-0.75]{}
\end{tikzpicture},\qquad x\in \Lambda_L,\quad\tau\in\frac{1}{2}\mathbb Z,
\ee
and different shades illustrate distinct matrices. 
 The function ${\rm mod}(x,n)$ indicates the remainder upon division by $n$.
Note that leftmost and rightmost gates are connected because of periodic boundary conditions. 

Finally, we point out that the dynamics generated by \eqref{eq:Floquet} are \emph{time-reversal invariant} if there exist a unitary operator $\mathbb K\in {\rm U}(d^{2L})$ such that~\cite{Mehtabook} 
\be
\mathbb K \mathbb U_{L}^{\phantom{T}} \mathbb K^\dag =  \mathbb U_{L}^T \qquad\text{and}\qquad  \mathbb K^{{T}}=\pm \mathbb K\,,
\label{eq:Tsym}
\ee
where $(\cdot)^T$ denotes transposition in the canonical basis \eqref{eq:realbasis} and $(\cdot)^\dagger$ Hermitian conjugation. Symmetric and antisymmetric matrices correspond respectively to cases where the anti-unitary operator implementing time reversal on the Hilbert space squares to plus or minus one~\cite{Mehtabook}. The first, ``regular'', kind of time-reversal symmetry emerges in physical systems with integer total angular momentum and is associated with orthogonal ensembles of RMT, while the second characterises systems with half-odd integer spin and is associated with symplectic ensembles~\cite{Mehtabook}.

\subsection{Our Setting}

Here we consider local gates of the form  
\begin{subequations}
\begin{align}
&U_{x+\frac{1}{2},\frac{1}{2}} = (u_{x} \otimes u_{x+\frac{1}{2}})\, U = 
\begin{tikzpicture}[baseline=(current  bounding  box.center), scale=.7]
\draw[ thick] (-4.5,0.75) -- (-3.25,-0.5);
\draw[ thick] (-4.25,-0.5) -- (-3,0.75);
\draw[ thick, fill=myred, rounded corners=2pt] (-4,0.25) rectangle (-3.5,-0.25);
\draw[thick] (-3.75,0.15) -- (-3.75+0.15,0.15) -- (-3.75+0.15,0);
\draw[ thick, fill=myblue3, rounded corners=2pt] (-3.25,.5) circle (.15);
\draw[ thick, fill=myblue4, rounded corners=2pt] (-4.25,.5) circle (.15);
\draw[thick] (-2*2.875+2.4,0.05+.5) -- (-2*2.875+2.55,0.05+.5) -- (-2*2.875+2.55,-0.1+.5);
\draw[thick]  (-2*2.875+1.45,-0.1+.5) -- (-2*2.875+1.45,0.05+.5) -- (-2*2.875+1.6,0.05+.5);
\Text[x=-4.25,y=-0.75]{}
\end{tikzpicture}\,,\label{eq:Floquetgates1}\\
&U_{x,1} =( w_{{\rm mod}(x-\frac{1}{2},L)} \otimes w_{x})\, W=\begin{tikzpicture}[baseline=(current  bounding  box.center), scale=.7]
\draw[ thick] (-4.5,0.75) -- (-3.25,-0.5);
\draw[ thick] (-4.25,-0.5) -- (-3,0.75);
\draw[ thick, fill=myorange, rounded corners=2pt] (-4,0.25) rectangle (-3.5,-0.25);
\draw[thick] (-3.75,0.15) -- (-3.75+0.15,0.15) -- (-3.75+0.15,0);
\draw[ thick, fill=myblue1, rounded corners=2pt] (-3.25,.5) circle (.15);
\draw[ thick, fill=myblue2, rounded corners=2pt] (-4.25,.5) circle (.15);
\draw[thick] (-2*2.875+2.4,0.05+.5) -- (-2*2.875+2.55,0.05+.5) -- (-2*2.875+2.55,-0.1+.5);
\draw[thick]  (-2*2.875+1.45,-0.1+.5) -- (-2*2.875+1.45,0.05+.5) -- (-2*2.875+1.6,0.05+.5);
\Text[x=-4.25,y=-0.75]{}
\end{tikzpicture}\,,\qquad  x\in \mathbb Z_{L},
\label{eq:Floquetgates2}
\end{align}
\end{subequations}
where $U,W\in {\rm U}(d^2)$ act non-trivially on a pair of neighbouring qudits and $u_{x},w_{x}\in {\rm U}(d)$ on a single one (we hence represented them graphically as balls acting on a single wire).
Therefore we have 
\begin{align}
&\mathbb U_{L} = 
\begin{tikzpicture}[baseline=(current  bounding  box.center), scale=0.55]
\draw[thick, dotted] (-9.5,-1.7) -- (0.4,-1.7);
\draw[thick, dotted] (-9.5,-1.3) -- (0.4,-1.3);
\foreach \i in {3,...,13}{
\draw[gray, dashed] (-12.5+\i,-2.1) -- (-12.5+\i,0.3);
}
\foreach \i in {3,...,5}{
\draw[gray, dashed] (-9.75,3-\i) -- (.75,3-\i);
}
\foreach \i in{0.5}
{
\draw[ thick] (0.5,1+2*\i-0.5-3.5) arc (-45:90:0.17);
\draw[ thick] (-10+0.5,1+2*\i-0.5-3.5) arc (270:90:0.15);
}
\foreach \i in{1.5}{
\draw[thick] (0.5,2*\i-0.5-3.5) arc (45:-90:0.17);
\draw[thick] (-10+0.5+0,2*\i-0.5-3.5) arc (90:270:0.15);
}
\foreach \i in {0,1}{
\Text[x=1.25,y=-2+2*\i]{\scriptsize$\i$}
}
\foreach \i in {1}{
\Text[x=1.25,y=-2+\i]{\small$\frac{\i}{2}$}
}
\foreach \i in {1,3,5}{
\Text[x=-7.5+\i+1-3,y=-2.6]{\small$\frac{\i}{2}$}
}
\foreach \i in {1,2,3}{
\Text[x=-7.5+2*\i-2,y=-2.6]{\scriptsize${\i}$}
}
\foreach \i in {0,...,4}{
\draw[thick] (-.5-2*\i,-1) -- (0.525-2*\i,0.025);
\draw[thick] (-0.525-2*\i,0.025) -- (0.5-2*\i,-1);
\draw[thick, fill=myorange, rounded corners=2pt] (-0.25-2*\i,-0.25) rectangle (.25-2*\i,-0.75);
\draw[thick] (-2*\i,-0.35) -- (-2*\i+0.15,-0.35) -- (-2*\i+0.15,-0.5);%
}
\foreach \jj[evaluate=\jj as \j using -2*(ceil(\jj/2)-\jj/2)] in {0}
\foreach \i in {1,...,5}
{
\draw[thick] (.5-2*\i-1*\j,-2-1*\jj) -- (1-2*\i-1*\j,-1.5-\jj);
\draw[thick] (1-2*\i-1*\j,-1.5-1*\jj) -- (1.5-2*\i-1*\j,-2-\jj);
\draw[thick] (.5-2*\i-1*\j,-1-1*\jj) -- (1-2*\i-1*\j,-1.5-\jj);
\draw[thick] (1-2*\i-1*\j,-1.5-1*\jj) -- (1.5-2*\i-1*\j,-1-\jj);
\draw[thick, fill=myred, rounded corners=2pt] (0.75-2*\i-1*\j,-1.75-\jj) rectangle (1.25-2*\i-1*\j,-1.25-\jj);
\draw[thick] (-2*\i+1,-1.35-\jj) -- (-2*\i+1.15,-1.35-\jj) -- (-2*\i+1.15,-1.5-\jj);%
}
\draw[ thick, fill=myblue1, rounded corners=2pt] (0.5,-1) circle (.15);
\draw[ thick, fill=myblue2, rounded corners=2pt] (-0.5,-1) circle (.15);
\draw[ thick, fill=myblue3, rounded corners=2pt] (-1.5,-1) circle (.15);
\draw[ thick, fill=myblue4, rounded corners=2pt] (-2.5,-1) circle (.15);
\draw[ thick, fill=myblue5, rounded corners=2pt] (-3.5,-1) circle (.15);
\draw[ thick, fill=myblue6, rounded corners=2pt] (-4.5,-1) circle (.15);
\draw[ thick, fill=myblue7, rounded corners=2pt] (-5.5,-1) circle (.15);
\draw[ thick, fill=myblue8, rounded corners=2pt] (-6.5,-1) circle (.15);
\draw[ thick, fill=myblue9, rounded corners=2pt] (-7.5,-1) circle (.15);
\draw[ thick, fill=myblue10, rounded corners=2pt] (-8.5,-1) circle (.15);
\draw[ thick, fill=myblue8, rounded corners=2pt] (0.5,0) circle (.15);
\draw[ thick, fill=myblue9, rounded corners=2pt] (-0.5,0) circle (.15);
\draw[ thick, fill=myblue1, rounded corners=2pt] (-1.5,0) circle (.15);
\draw[ thick, fill=myblue3, rounded corners=2pt] (-2.5,0) circle (.15);
\draw[ thick, fill=myblue2, rounded corners=2pt] (-3.5,0) circle (.15);
\draw[ thick, fill=myblue4, rounded corners=2pt] (-4.5,0) circle (.15);
\draw[ thick, fill=myblue10, rounded corners=2pt] (-5.5,0) circle (.15);
\draw[ thick, fill=myblue9, rounded corners=2pt] (-6.5,0) circle (.15);
\draw[ thick, fill=myblue8, rounded corners=2pt] (-7.5,0) circle (.15);
\draw[ thick, fill=myblue7, rounded corners=2pt] (-8.5,0) circle (.15);
\foreach \jj[evaluate=\jj as \j using -2*(ceil(\jj/2)-\jj/2)] in {0}
\foreach \i in {1,...,5}{
\draw[thick] (-2*\i+1.4,-.95-\jj) -- (-2*\i+1.55,-.95-\jj) -- (-2*\i+1.55,-1.1-\jj);
}
\foreach \jj[evaluate=\jj as \j using -2*(ceil(\jj/2)-\jj/2)] in {0}
\foreach \i in {1,...,5}{
\draw[thick] (-2*\i+2.4,0.05-\jj) -- (-2*\i+2.55,0.05-\jj) -- (-2*\i+2.55,-0.1-\jj);
}
\foreach \jj[evaluate=\jj as \j using -2*(ceil(\jj/2)-\jj/2)] in {0}
\foreach \i in {1,...,5}{
\draw[thick]  (-2*\i+1.45,-0.1-\jj) -- (-2*\i+1.45,0.05-\jj) -- (-2*\i+1.6,0.05-\jj);
}
\foreach \jj[evaluate=\jj as \j using -2*(ceil(\jj/2)-\jj/2)] in {0}
\foreach \i in {0,...,4}{
\draw[thick]  (-2*\i+.45,-1.1-\jj) -- (-2*\i+.45,-0.95-\jj) -- (-2*\i+.6,-0.95-\jj);
}
\Text[x=-2,y=-2.6]{$\cdots$}
\end{tikzpicture}.
\label{eq:diagramFloquet}
\end{align}
In particular, it is immediate to see that, choosing local gates \eqref{eq:Floquetgates1}--\eqref{eq:Floquetgates2} with    
\be
U=U^T, \qquad W=W^T, \qquad w_{x}^{\phantom{T}}=u^{T}_{x},\qquad \forall\,\, x\in \Lambda_L,
\label{eq:timereversal}
\ee
the condition \eqref{eq:Tsym} is fulfilled with  
\be
\mathbb K=\prod_{x \in \mathbb Z_{L}+\frac{1}{2}} \eta_{x,L}(W)= \mathbb K^T \,.
\ee
Namely, the dynamics generated by \eqref{eq:diagramFloquet} are time-reversal invariant. Here we consider both the time-reversal-invariant and the non-time-reversal-invariant cases.

We remark that in \eqref{eq:Floquetgates1}--\eqref{eq:Floquetgates2} we assumed the 2-site gates $U,W$ to be the same for all $x$. In physical terms this means that we consider interactions that are \emph{homogeneous} in space, while we allow for some position-dependent `external fields' (encoded in the single-site gates $u_{x}, w_x$). The extension of our results to fully inhomogeneous systems is discussed in Sec.~\ref{Sec:inhom}.

\subsection{Spectral form factor}
\label{sec:SFF}

The objective of this paper is the study of spectral statistics of the Floquet operator. Namely, we consider the distribution of the elements of the spectrum of the unitary Floquet matrix
\be
{\rm spect}[\mathbb U_{L}]= \{e^{i \varphi_j};\,j=1,2\ldots,{\cal N}\}\,,
\ee
where $\varphi_j$ --- conventionally referred to as \emph{quasienergies} --- can be taken to be in $[0,2\pi)$. Considering ${\rm spect}[\mathbb U_{L}]$ as a one-dimensional gas on the circle ${\cal S}^1$, 
we analyse its 2-point correlation functions, specifically,  the \emph{spectral form factor} (SFF) defined as 
\be
K(t,L) := \mathbb E\!\left[\,|{\rm tr}\, \mathbb U_{L}^t|^2\,\right] = \mathbb E\left[\sum_{j,j'=1}^{\mathcal N} e^{i (\varphi_j-\varphi_{j'}) t}\right], \qquad t,L\in\mathbb N\,.
\label{eq:SFF}
\ee
Here $\mathbb E[\cdot]$ is an average over an ensemble of similar systems. The average is necessary to smear out the fluctuations of 
$|{\rm tr}\, \mathbb U_{L}^t|^2$, which do not die out even in the limit of large $L$, and extract the universal behaviour. We shall see later that very mild forms of averaging are sufficient (we remind the reader that the results are most interesting in the limit of clean systems), 
specifically we will consider cases where $u_x$ and $w_x$ are \iid for $x\in\Lambda_L$ densely covering an arbitrary small ball around the identity in ${\rm SU}(d)$.
The SFF is directly connected to the Fourier transform of the quasi-energy 2-point function. Indeed, introducing the $n$-point function as
\be
\rho_{n}(\vartheta_1,\ldots,\vartheta_n) := 
\mathbb E\left[  \sum_{j_1\neq\ldots \neq j_n=1}^{\mathcal N} \prod_{k=1}^n \!\delta(\vartheta_k-\varphi_{j_k})\!\right],
\ee
the SFF can be expressed as 
\be
K(t,L) =  \int\limits_{[0,2\pi]^{2}} \!\!{\rm d} \vartheta_1 {\rm d} \vartheta_2\,\,  e^{i (\vartheta_1- \vartheta_2) t}  \rho_{2}(\vartheta_1, \vartheta_2)+ \mathcal N.
\label{eq:SFF2pointf}
\ee
Since this quantity measures correlations between quasienergy levels at arbitrary distance, it is very convenient to analyse extended systems for large volume $L$ where neighbouring quasienergy levels become exponentially close in $L$ and one has to look at correlations on larger scales. 

\subsection{Spectral form factor for random unitary matrices} 
\label{sec:RMT}
Before moving to the analysis of \eqref{eq:SFF} for local quantum circuits let us briefly recall our point of reference: the SFF of random unitary matrices. In this case the average $\mathbb E\left[\cdot\right]$ in \eqref{eq:SFF} is replaced by the integration over an ensemble of random unitary matrices of dimension ${\cal N}$, i.e. 
\be
K_{\rm ens}(t,{\cal N}) := \int |{\rm tr}\, \mathbb U^t|^2 {\rm d}\mu_{\rm ens}(\mathbb U),
\ee
and the result depends on the precise form of the measure ${\rm d}\mu_{\rm ens}(\mathbb U)$ of the ensemble considered. Specifically, in this paper we are interested in the two most common cases:
(i) systems without anti-unitary symmetries, and (ii) systems with regular time-reversal symmetry (squaring to the identity).
In these two cases the relevant ensembles of random matrices are two of Dyson's circular ensembles: the {\em Circular Unitary Ensemble} (CUE) and the {\em Circular Orthogonal Ensemble} (COE). The CUE measure is the invariant Haar measure over ${\rm U}(\mathcal N)$, while the COE measure is defined for  \emph{symmetric unitary matrices} and is uniquely specified by the property of being invariant under orthogonal transformations~\cite{Mehtabook}. In these two cases the result reads as~\cite{Haakebook} 
\begin{align}
K_{\rm CUE}(t,{\cal N}) &= {\rm min}(t, \mathcal N)\,,\\
K_{\rm COE}(t,{\cal N}) &= 2{\rm min}(t, \mathcal N)\left(1-\sum_{m=1}^{{\rm min}(t, \mathcal N)} \frac{1}{2m+2{\rm max}(t,\mathcal N)-\mathcal N-1}\right)\!.
\end{align}
In particular, in the thermodynamic limit $L\to\infty$ (${\cal N}\to\infty$) they simplify to
\be
\lim_{{\cal N}\to\infty} K_{\rm CUE}(t,{\cal N}) = t,\qquad\qquad \lim_{{\cal N}\to\infty} K_{\rm COE}(t,{\cal N}) = 2 t\,.
\label{eq:SFFRMTTL}
\ee
The main result of our paper is the proof that one recovers the r.h.s.\ of \eqref{eq:SFFRMTTL} by computing exactly the expression \eqref{eq:SFF} for a broad class of Floquet quantum circuits.

\subsection{Spectral form factor of Floquet quantum circuits}

For local quantum circuits the SFF \eqref{eq:SFF} can be represented diagrammatically as follows 
\be
K(t,L) = 
\mathbb E\left[ {\rm tr}(\mathbb U_{L})^t {\rm tr}(\mathbb U_{L}^\dag)^t\right]=
 \mathbb E\Biggl[\begin{tikzpicture}[baseline=(current  bounding  box.center), scale=0.49]
\foreach \i in {1,...,5}{
\draw[thick, dotted] (2*\i+2-12.5+0.255,-1.75-0.1) -- (2*\i+2-12.5+0.255,4.25-0.1);
\draw[thick, dotted] (2*\i+2-11.5-0.255,-1.75-0.1) -- (2*\i+2-11.5-0.255,4.25-0.1);}

\foreach \i in {1,...,5}{
\draw[thick] (2*\i+2-11.5,4) arc (-45:175:0.15);
\draw[thick] (2*\i+2-11.5,-2) arc (315:180:0.15);
\draw[thick] (2*\i+2-0.5-12,-2) arc (-135:0:0.15);
}
\foreach \i in {2,...,6}{
\draw[thick] (2*\i+2-2.5-12,4) arc (225:0:0.15);
}
\foreach \i in {0,1,2}{
\draw[thick, dotted] (-9.5,2*\i-1.745) -- (0.4,2*\i-1.745);
\draw[thick, dotted] (-9.5,2*\i-1.255) -- (0.4,2*\i-1.255);
}
\foreach \i in{1.5,2.5,3.5}{
\draw[thick] (0.5,2*\i-0.5-3.5) arc (45:-90:0.15);
\draw[thick] (-10+0.5+0,2*\i-0.5-3.5) arc (45:270:0.15);
}
\foreach \i in{0.5,1.5,2.5}
{
\draw[ thick] (0.5,1+2*\i-0.5-3.5) arc (-45:90:0.15);
\draw[ thick] (-10+0.5,1+2*\i-0.5-3.5) arc (315:90:0.15);
}
\foreach \jj[evaluate=\jj as \j using -2*(ceil(\jj/2)-\jj/2)] in {-1,-3,-5}{
\foreach \i in {1,...,5}
{
\draw[thick] (.5-2*\i-1*\j,-2-1*\jj) -- (1-2*\i-1*\j,-1.5-\jj);
\draw[thick] (1-2*\i-1*\j,-1.5-1*\jj) -- (1.5-2*\i-1*\j,-2-\jj);
\draw[thick] (.5-2*\i-1*\j,-1-1*\jj) -- (1-2*\i-1*\j,-1.5-\jj);
\draw[thick] (1-2*\i-1*\j,-1.5-1*\jj) -- (1.5-2*\i-1*\j,-1-\jj);
\draw[thick, fill=myorange, rounded corners=2pt] (0.75-2*\i-1*\j,-1.75-\jj) rectangle (1.25-2*\i-1*\j,-1.25-\jj);
\draw[thick] (-2*\i+2,-1.35-\jj) -- (-2*\i+2.15,-1.35-\jj) -- (-2*\i+2.15,-1.5-\jj);%
}
}
\foreach \jj[evaluate=\jj as \j using -2*(ceil(\jj/2)-\jj/2)] in {-4,-2,0}{
\foreach \i in {1,...,5}
{
\draw[thick] (.5-2*\i-1*\j,-2-1*\jj) -- (1-2*\i-1*\j,-1.5-\jj);
\draw[thick] (1-2*\i-1*\j,-1.5-1*\jj) -- (1.5-2*\i-1*\j,-2-\jj);
\draw[thick] (.5-2*\i-1*\j,-1-1*\jj) -- (1-2*\i-1*\j,-1.5-\jj);
\draw[thick] (1-2*\i-1*\j,-1.5-1*\jj) -- (1.5-2*\i-1*\j,-1-\jj);
\draw[thick, fill=myred, rounded corners=2pt] (0.75-2*\i-1*\j,-1.75-\jj) rectangle (1.25-2*\i-1*\j,-1.25-\jj);
\draw[thick] (-2*\i+1,-1.35-\jj) -- (-2*\i+1.15,-1.35-\jj) -- (-2*\i+1.15,-1.5-\jj);%
}
}
\foreach \jj in {0,2,4}{
\draw[ thick, fill=myblue1, rounded corners=2pt] (0.5,-1+\jj) circle (.15);
\draw[ thick, fill=myblue2, rounded corners=2pt] (-0.5,-1+\jj) circle (.15);
\draw[ thick, fill=myblue3, rounded corners=2pt] (-1.5,-1+\jj) circle (.15);
\draw[ thick, fill=myblue4, rounded corners=2pt] (-2.5,-1+\jj) circle (.15);
\draw[ thick, fill=myblue5, rounded corners=2pt] (-3.5,-1+\jj) circle (.15);
\draw[ thick, fill=myblue6, rounded corners=2pt] (-4.5,-1+\jj) circle (.15);
\draw[ thick, fill=myblue7, rounded corners=2pt] (-5.5,-1+\jj) circle (.15);
\draw[ thick, fill=myblue8, rounded corners=2pt] (-6.5,-1+\jj) circle (.15);
\draw[ thick, fill=myblue9, rounded corners=2pt] (-7.5,-1+\jj) circle (.15);
\draw[ thick, fill=myblue10, rounded corners=2pt] (-8.5,-1+\jj) circle (.15);
\draw[ thick, fill=myblue8, rounded corners=2pt] (0.5,\jj) circle (.15);
\draw[ thick, fill=myblue9, rounded corners=2pt] (-0.5,\jj) circle (.15);
\draw[ thick, fill=myblue1, rounded corners=2pt] (-1.5,\jj) circle (.15);
\draw[ thick, fill=myblue3, rounded corners=2pt] (-2.5,\jj) circle (.15);
\draw[ thick, fill=myblue2, rounded corners=2pt] (-3.5,\jj) circle (.15);
\draw[ thick, fill=myblue4, rounded corners=2pt] (-4.5,\jj) circle (.15);
\draw[ thick, fill=myblue10, rounded corners=2pt] (-5.5,\jj) circle (.15);
\draw[ thick, fill=myblue9, rounded corners=2pt] (-6.5,\jj) circle (.15);
\draw[ thick, fill=myblue8, rounded corners=2pt] (-7.5,\jj) circle (.15);
\draw[ thick, fill=myblue7, rounded corners=2pt] (-8.5,\jj) circle (.15);
}
\foreach \jj in {0,-2,-4}{
\foreach \i in {1,...,5}{
\draw[thick] (-2*\i+1.4,-.95-\jj) -- (-2*\i+1.55,-.95-\jj) -- (-2*\i+1.55,-1.1-\jj);}
\foreach \i in {1,...,5}{
\draw[thick] (-2*\i+2.4,0.05-\jj) -- (-2*\i+2.55,0.05-\jj) -- (-2*\i+2.55,-0.1-\jj);}
\foreach \i in {1,...,5}{
\draw[thick]  (-2*\i+1.45,-0.1-\jj) -- (-2*\i+1.45,0.05-\jj) -- (-2*\i+1.6,0.05-\jj);}
\foreach \i in {0,...,4}{
\draw[thick]  (-2*\i+.45,-1.1-\jj) -- (-2*\i+.45,-0.95-\jj) -- (-2*\i+.6,-0.95-\jj);}
}
\def\shiftx{0}
\def\shifty{-9}
\foreach \i in {1,...,5}{
\draw[ thick, dotted] (2*\i+2-1.485+0.25+\shiftx-11,-2.5-0.1+\shifty+2.5) -- (2*\i+2-1.485+0.25+\shiftx-11,3.5-0.1+\shifty+2.5);
\draw[ thick, dotted] (2*\i+2-0.525-0.25+\shiftx-11,-2.5-0.1+\shifty+2.5) -- (2*\i+2-0.525-0.25+\shiftx-11,3.5-0.1+\shifty+2.5);
}
\foreach \i in {0,1,2}{
\draw[ thick, dotted] (1.75+\shiftx-11,2*\i-1.25+\shifty+2.5) -- (11.5+\shiftx-11,2*\i-1.25+\shifty+2.5);
\draw[ thick, dotted] (1.5+\shiftx-11,2*\i-.76+\shifty+2.5) -- (11.5+\shiftx-11,2*\i-.76+\shifty+2.5);
}
\foreach \i in {1,...,5}
{
\draw[ thick] (2*\i+2-1.5-11+\shiftx,3.5+\shifty+2.5) arc (135:-0:0.15);
\draw[ thick] (2*\i+2-.5-11+\shiftx,3.5+\shifty+2.5) arc (-325:-180:0.15);
\draw[ thick] (2*\i+2-1.5-11+\shiftx,-2.5+\shifty+2.5) arc (-45:180:-0.15);
\draw[ thick] (2*\i+2-0.5-11+\shiftx,-2.5+\shifty+2.5) arc (45:-180:0.15);
}
\foreach \i in {3,...,5}
{
\draw[ thick] (\shiftx+.5,2*\i-0.5-3.5+\shifty) arc (45:-90:0.15);
\draw[ thick] (\shiftx-10+0.5+0,2*\i-0.5-3.5+\shifty) arc (45:280:0.15);
}
\foreach \i in{2,...,4}
{
\draw[ thick] (\shiftx+.5,1+2*\i-0.5-3.5+\shifty) arc (-45:90:0.15);
\draw[ thick] (\shiftx-10+0.5,1+2*\i-0.5-3.5+\shifty) arc (-45:-280:0.15);
}
\foreach \i in {1,...,5}
{
\draw[ thick] (\shiftx+.5-2*\i,6+\shifty) -- (\shiftx+1-2*\i,5.5+\shifty);
\draw[ thick] (\shiftx+1.5-2*\i,6+\shifty) -- (\shiftx+1-2*\i,5.5+\shifty);
}
\foreach \jj[evaluate=\jj as \j using -2*(ceil(\jj/2)-\jj/2)] in {0,...,3}
\foreach \i in {1,...,5}
{
\draw[ thick] (\shiftx+.5-2*\i-1*\j,2+1*\jj+\shifty) -- (\shiftx+1-2*\i-1*\j,1.5+\jj+\shifty);
\draw[ thick] (\shiftx+1-2*\i-1*\j,1.5+1*\jj+\shifty) -- (\shiftx+1.5-2*\i-1*\j,2+\jj+\shifty);
}
\foreach \i in {0,...,4}
{
\draw[ thick] (\shiftx-.5-2*\i,1+\shifty) -- (\shiftx+0.5-2*\i,0+\shifty);
\draw[ thick] (\shiftx-0.5-2*\i,0+\shifty) -- (\shiftx+0.5-2*\i,1+\shifty);
\draw[ thick, fill=mygreen, rounded corners=2pt] (\shiftx-0.25-2*\i,0.25+\shifty) rectangle (\shiftx+.25-2*\i,0.75+\shifty);
\draw[thick] (\shiftx-2*\i,0.65+\shifty) -- (\shiftx+.15-2*\i,.65+\shifty) -- (\shiftx+.15-2*\i,0.5+\shifty);
}
\foreach \jj[evaluate=\jj as \j using -2*(ceil(\jj/2)-\jj/2)] in {-1,1,3}
\foreach \i in {1,...,5}
{
\draw[ thick] (\shiftx+.5-2*\i-1*\j,1+1*\jj+\shifty) -- (\shiftx+1-2*\i-1*\j,1.5+\jj+\shifty);
\draw[ thick] (\shiftx+1-2*\i-1*\j,1.5+1*\jj+\shifty) -- (\shiftx+1.5-2*\i-1*\j,1+\jj+\shifty);
\draw[ thick, fill=myblue4, rounded corners=2pt] (\shiftx+0.75-2*\i-1*\j,1.75+\jj+\shifty) rectangle (\shiftx+1.25-2*\i-1*\j,1.25+\jj+\shifty);
\draw[thick] (\shiftx+1-2*\i-1*\j,1.65+1*\jj+\shifty) -- (\shiftx+1.15-2*\i-1*\j,1.65+1*\jj+\shifty) -- (\shiftx+1.15-2*\i-1*\j,1.5+1*\jj+\shifty);
}
\foreach \jj[evaluate=\jj as \j using -2*(ceil(\jj/2)-\jj/2)] in {0,2,4}{
\foreach \i in {1,...,5}
{
\draw[ thick] (\shiftx+.5-2*\i-1*\j,1+1*\jj+\shifty) -- (\shiftx+1-2*\i-1*\j,1.5+\jj+\shifty);
\draw[ thick] (\shiftx+1-2*\i-1*\j,1.5+1*\jj+\shifty) -- (\shiftx+1.5-2*\i-1*\j,1+\jj+\shifty);
\draw[ thick, fill=myblue, rounded corners=2pt] (\shiftx+0.75-2*\i-1*\j,1.75+\jj+\shifty) rectangle (\shiftx+1.25-2*\i-1*\j,1.25+\jj+\shifty);
\draw[thick] (\shiftx+1-2*\i-1*\j,1.65+1*\jj+\shifty) -- (\shiftx+1.15-2*\i-1*\j,1.65+1*\jj+\shifty) -- (\shiftx+1.15-2*\i-1*\j,1.5+1*\jj+\shifty);
}}
\foreach \jj in {0.5,2.5,4.5}{
\draw[ thick, fill=mygray1, rounded corners=2pt] (.5+\shiftx,-.5+\jj+\shifty) circle (.15);
\draw[ thick, fill=mygray4, rounded corners=2pt] (-.5+\shiftx,-.5+\jj+\shifty) circle (.15);
\draw[ thick, fill=mygray3, rounded corners=2pt] (-1.5+\shiftx,-.5+\jj+\shifty) circle (.15);
\draw[ thick, fill=mygray2, rounded corners=2pt] (-2.5+\shiftx,-.5+\jj+\shifty) circle (.15);
\draw[ thick, fill=mygray1, rounded corners=2pt] (-3.5+\shiftx,-.5+\jj+\shifty) circle (.15);
\draw[ thick, fill=mygray5, rounded corners=2pt] (-4.5+\shiftx,-.5+\jj+\shifty) circle (.15);
\draw[ thick, fill=mygray4, rounded corners=2pt] (-5.5+\shiftx,-.5+\jj+\shifty) circle (.15);
\draw[ thick, fill=mygray8, rounded corners=2pt] (-6.5+\shiftx,-.5+\jj+\shifty) circle (.15);
\draw[ thick, fill=mygray7, rounded corners=2pt] (-7.5+\shiftx,-.5+\jj+\shifty) circle (.15);
\draw[ thick, fill=mygray6, rounded corners=2pt] (-8.5+\shiftx,-.5+\jj+\shifty) circle (.15);
}
\foreach \jj in {1.5,3.5,5.5}
{
\draw[ thick, fill=mygray3, rounded corners=2pt] (.5+\shiftx,-.5+\jj+\shifty) circle (.15);
\draw[ thick, fill=mygray6, rounded corners=2pt] (-.5+\shiftx,-.5+\jj+\shifty) circle (.15);
\draw[ thick, fill=mygray1, rounded corners=2pt] (-1.5+\shiftx,-.5+\jj+\shifty) circle (.15);
\draw[ thick, fill=mygray2, rounded corners=2pt] (-2.5+\shiftx,-.5+\jj+\shifty) circle (.15);
\draw[ thick, fill=mygray3, rounded corners=2pt] (-3.5+\shiftx,-.5+\jj+\shifty) circle (.15);
\draw[ thick, fill=mygray4, rounded corners=2pt] (-4.5+\shiftx,-.5+\jj+\shifty) circle (.15);
\draw[ thick, fill=mygray5, rounded corners=2pt] (-5.5+\shiftx,-.5+\jj+\shifty) circle (.15);
\draw[ thick, fill=mygray6, rounded corners=2pt] (-6.5+\shiftx,-.5+\jj+\shifty) circle (.15);
\draw[ thick, fill=mygray7, rounded corners=2pt] (-7.5+\shiftx,-.5+\jj+\shifty) circle (.15);
\draw[ thick, fill=mygray8, rounded corners=2pt] (-8.5+\shiftx,-.5+\jj+\shifty) circle (.15);
}
\foreach \jj in {0,-2,-4}{
\foreach \i in {1,...,5}{
\draw[thick] (-2*\i+1.4+\shiftx,-.95-\jj-0.5+\shifty+1.5) -- (-2*\i+1.55+\shiftx,-.95-\jj-0.5+\shifty+1.5) -- (-2*\i+1.55+\shiftx,-1.1-\jj-0.5+\shifty+1.5);}
\foreach \i in {1,...,5}{
\draw[thick] (-2*\i+2.4+\shiftx,0.05-\jj-0.5+\shifty+1.5) -- (-2*\i+2.55+\shiftx,0.05-\jj-0.5+\shifty+1.5) -- (-2*\i+2.55+\shiftx,-0.1-\jj-0.5+\shifty+1.5);}
\foreach \i in {1,...,5}{
\draw[thick]  (-2*\i+1.45+\shiftx,-0.1-\jj-0.5+\shifty+1.5) -- (-2*\i+1.45+\shiftx,0.05-\jj-0.5+\shifty+1.5) -- (-2*\i+1.6+\shiftx,0.05-\jj-0.5+\shifty+1.5);}
\foreach \i in {0,...,4}{
\draw[thick]  (-2*\i+.45+\shiftx,-1.1-\jj-0.5+\shifty+1.5) -- (-2*\i+.45+\shiftx,-0.95-\jj-0.5+\shifty+1.5) -- (-2*\i+.6+\shiftx,-0.95-\jj-0.5+\shifty+1.5);}
}
\end{tikzpicture}\Biggr],
\ee
where we represented the trace in the {\em forward time sheet} (${\rm tr}\, \mathbb U_{L}^t$) using the diagram \eqref{eq:diagramFloquet} and that in the {\em backward time sheet} (${\rm tr}\, (\mathbb U_{L}^\dag)^t$) by introducing  
\begin{align}
 U^\dag & = 
\begin{tikzpicture}[baseline=(current  bounding  box.center), scale=.7]
\draw[ thick] (-4.25,0.5) -- (-3.25,-0.5);
\draw[ thick] (-4.25,-0.5) -- (-3.25,0.5);
\draw[ thick, fill=myblue, rounded corners=2pt] (-4,0.25) rectangle (-3.5,-0.25);
\draw[thick] (-3.75,0.15) -- (-3.75+0.15,0.15) -- (-3.75+0.15,0);
\Text[x=-4.25,y=-0.75]{}
\end{tikzpicture}\,,
& W^\dag &=\begin{tikzpicture}[baseline=(current  bounding  box.center), scale=.7]
\draw[ thick] (-4.25,0.5) -- (-3.25,-0.5);
\draw[ thick] (-4.25,-0.5) -- (-3.25,0.5);
\draw[ thick, fill=myblue4, rounded corners=2pt] (-4,0.25) rectangle (-3.5,-0.25);
\draw[thick] (-3.75,0.15) -- (-3.75+0.15,0.15) -- (-3.75+0.15,0);
\Text[x=-4.25,y=-0.75]{}
\end{tikzpicture}\,,
&
u_{x}^\dag, w_{x}^\dag=
\begin{tikzpicture}[baseline=(current  bounding  box.center), scale=.7]
\draw[ thick] (-4.25,0.5) -- (-4.25,-0.5);
\draw[ thick, fill=mygray4, rounded corners=2pt] (-4.25,0) circle (.15);
\draw[thick, rotate around = {-45:(0.525-4.77,0.375-0.4)}]  (.45-4.77,0.3-0.4) -- (.45-4.77,0.45-0.4) -- (.6-4.77,0.45-0.4);
\Text[x=-4.25,y=-0.75]{}
\end{tikzpicture}\,.
\end{align}
Once again shades of the same colour denote different matrices. Note that top and bottom lines at the same positions within both sheets are connected because of the traces. 

Folding the backward sheet (blue) underneath the forward one (red) we write the folded circuit representation of the SFF   
\be
K(t,L)= \mathbb{E}\Biggl[
\begin{tikzpicture}[baseline=(current  bounding  box.center), scale=0.55]
\foreach \i in {1,...,5}{
\draw[very thick, dotted] (2*\i+2-12.5+0.255,-1.75-0.1) -- (2*\i+2-12.5+0.255,4.25-0.1);
\draw[very thick, dotted] (2*\i+2-11.5-0.255,-1.75-0.1) -- (2*\i+2-11.5-0.255,4.25-0.1);}

\foreach \i in {1,...,5}{
\draw[very thick] (2*\i+2-11.5,4) arc (-45:175:0.15);
\draw[very thick] (2*\i+2-11.5,-2) arc (315:180:0.15);
\draw[very thick] (2*\i+2-0.5-12,-2) arc (-135:0:0.15);
}
\foreach \i in {2,...,6}{
\draw[very thick] (2*\i+2-2.5-12,4) arc (225:0:0.15);
}
\foreach \i in {0,1,2}{
\draw[very thick, dotted] (-9.5,2*\i-1.745) -- (0.4,2*\i-1.745);
\draw[very thick, dotted] (-9.5,2*\i-1.255) -- (0.4,2*\i-1.255);
}
\foreach \i in{1.5,2.5,3.5}{
\draw[very thick] (0.5,2*\i-0.5-3.5) arc (45:-90:0.15);
\draw[very thick] (-10+0.5+0,2*\i-0.5-3.5) arc (45:270:0.15);
}
\foreach \i in{0.5,1.5,2.5}
{
\draw[very thick] (0.5,1+2*\i-0.5-3.5) arc (-45:90:0.15);
\draw[very thick] (-10+0.5,1+2*\i-0.5-3.5) arc (315:90:0.15);
}
\foreach \jj[evaluate=\jj as \j using -2*(ceil(\jj/2)-\jj/2)] in {-1,-3,-5}{
\foreach \i in {1,...,5}
{
\draw[very thick] (.5-2*\i-1*\j,-2-1*\jj) -- (1-2*\i-1*\j,-1.5-\jj);
\draw[very thick] (1-2*\i-1*\j,-1.5-1*\jj) -- (1.5-2*\i-1*\j,-2-\jj);
\draw[very thick] (.5-2*\i-1*\j,-1-1*\jj) -- (1-2*\i-1*\j,-1.5-\jj);
\draw[very thick] (1-2*\i-1*\j,-1.5-1*\jj) -- (1.5-2*\i-1*\j,-1-\jj);
\draw[thick, fill=OliveGreen, rounded corners=2pt] (0.75-2*\i-1*\j,-1.75-\jj) rectangle (1.25-2*\i-1*\j,-1.25-\jj);
\draw[ thick] (-2*\i+2,-1.35-\jj) -- (-2*\i+2.15,-1.35-\jj) -- (-2*\i+2.15,-1.5-\jj);%
}
}
\foreach \jj[evaluate=\jj as \j using -2*(ceil(\jj/2)-\jj/2)] in {-4,-2,0}{
\foreach \i in {1,...,5}
{
\draw[very thick] (.5-2*\i-1*\j,-2-1*\jj) -- (1-2*\i-1*\j,-1.5-\jj);
\draw[very thick] (1-2*\i-1*\j,-1.5-1*\jj) -- (1.5-2*\i-1*\j,-2-\jj);
\draw[very thick] (.5-2*\i-1*\j,-1-1*\jj) -- (1-2*\i-1*\j,-1.5-\jj);
\draw[very thick] (1-2*\i-1*\j,-1.5-1*\jj) -- (1.5-2*\i-1*\j,-1-\jj);
\draw[thick, fill=mygreen, rounded corners=2pt] (0.75-2*\i-1*\j,-1.75-\jj) rectangle (1.25-2*\i-1*\j,-1.25-\jj);
\draw[ thick] (-2*\i+1,-1.35-\jj) -- (-2*\i+1.15,-1.35-\jj) -- (-2*\i+1.15,-1.5-\jj);%
}
}
\foreach \jj in {0,2,4}{
\draw[ thick, fill=myyellow1, rounded corners=2pt] (0.5,-1+\jj) circle (.15);
\draw[ thick, fill=myyellow2, rounded corners=2pt] (-0.5,-1+\jj) circle (.15);
\draw[ thick, fill=myyellow3, rounded corners=2pt] (-1.5,-1+\jj) circle (.15);
\draw[ thick, fill=myyellow4, rounded corners=2pt] (-2.5,-1+\jj) circle (.15);
\draw[ thick, fill=myyellow5, rounded corners=2pt] (-3.5,-1+\jj) circle (.15);
\draw[ thick, fill=myyellow6, rounded corners=2pt] (-4.5,-1+\jj) circle (.15);
\draw[ thick, fill=myyellow7, rounded corners=2pt] (-5.5,-1+\jj) circle (.15);
\draw[ thick, fill=myyellow8, rounded corners=2pt] (-6.5,-1+\jj) circle (.15);
\draw[ thick, fill=myyellow9, rounded corners=2pt] (-7.5,-1+\jj) circle (.15);
\draw[ thick, fill=myyellow10, rounded corners=2pt] (-8.5,-1+\jj) circle (.15);
\draw[ thick, fill=myyellow8, rounded corners=2pt] (0.5,\jj) circle (.15);
\draw[ thick, fill=myyellow9, rounded corners=2pt] (-0.5,\jj) circle (.15);
\draw[ thick, fill=myyellow1, rounded corners=2pt] (-1.5,\jj) circle (.15);
\draw[ thick, fill=myyellow3, rounded corners=2pt] (-2.5,\jj) circle (.15);
\draw[ thick, fill=myyellow2, rounded corners=2pt] (-3.5,\jj) circle (.15);
\draw[ thick, fill=myyellow4, rounded corners=2pt] (-4.5,\jj) circle (.15);
\draw[ thick, fill=myyellow10, rounded corners=2pt] (-5.5,\jj) circle (.15);
\draw[ thick, fill=myyellow9, rounded corners=2pt] (-6.5,\jj) circle (.15);
\draw[ thick, fill=myyellow8, rounded corners=2pt] (-7.5,\jj) circle (.15);
\draw[ thick, fill=myyellow7, rounded corners=2pt] (-8.5,\jj) circle (.15);
}
\foreach \jj in {0,-2,-4}{
\foreach \i in {1,...,5}{
\draw[ thick] (-2*\i+1.4,-.95-\jj) -- (-2*\i+1.55,-.95-\jj) -- (-2*\i+1.55,-1.1-\jj);}
\foreach \i in {1,...,5}{
\draw[ thick] (-2*\i+2.4,0.05-\jj) -- (-2*\i+2.55,0.05-\jj) -- (-2*\i+2.55,-0.1-\jj);}
\foreach \i in {1,...,5}{
\draw[ thick]  (-2*\i+1.45,-0.1-\jj) -- (-2*\i+1.45,0.05-\jj) -- (-2*\i+1.6,0.05-\jj);}
\foreach \i in {0,...,4}{
\draw[ thick]  (-2*\i+.45,-1.1-\jj) -- (-2*\i+.45,-0.95-\jj) -- (-2*\i+.6,-0.95-\jj);}
}
\end{tikzpicture}\Biggr]\,,
\label{eq:SFFfolded}
\ee
where we introduced ``doubled'' or thickened wires
\be
\label{eq:thickwire}
\begin{tikzpicture}[baseline=(current  bounding  box.center), scale=0.8]
\def\eps{0.5};
\draw[very thick] (-3.25,0.5) -- (-3.25,-0.5);
\draw[ thick] (-2.0,0.5) -- (-2.0,-0.5);
\draw[ thick] (-1.75,0.7) -- (-1.75,-0.3);
\Text[x=-2.65,y=0.05]{$=$}
\end{tikzpicture}\,,
\ee
and ``doubled'' gates
\begin{eqnarray}
\label{eq:doublegate}
&\begin{tikzpicture}[baseline=(current  bounding  box.center), scale=.7]
\def\eps{0.5};
\Wgategreen{-3.75}{0};
\Text[x=-3.75,y=-0.6]{}
\end{tikzpicture}
=
\begin{tikzpicture}[baseline=(current  bounding  box.center), scale=.7]
\draw[thick] (-1.65,0.65) -- (-0.65,-0.35);
\draw[thick] (-1.65,-0.35) -- (-0.65,0.65);
\draw[ thick, fill=myblue, rounded corners=2pt] (-1.4,0.4) rectangle (-.9,-0.1);
\draw[thick] (-1.15,0) -- (-1,0) -- (-1,0.15);
\draw[thick] (-2.25,0.5) -- (-1.25,-0.5);
\draw[thick] (-2.25,-0.5) -- (-1.25,0.5);
\draw[ thick, fill=myred, rounded corners=2pt] (-2,0.25) rectangle (-1.5,-0.25);
\draw[thick] (-1.75,0.15) -- (-1.6,0.15) -- (-1.6,0);
\Text[x=-2.25,y=-0.6]{}
\end{tikzpicture}
= U\otimes U^{*}\,,
\qquad
\begin{tikzpicture}[baseline=(current  bounding  box.center), scale=.7]
\def\eps{0.5};
\Wgateolivegreen{-3.75}{0};
\Text[x=-3.75,y=-0.6]{}
\end{tikzpicture}
=
\begin{tikzpicture}[baseline=(current  bounding  box.center), scale=.7]
\draw[thick] (-1.65,0.65) -- (-0.65,-0.35);
\draw[thick] (-1.65,-0.35) -- (-0.65,0.65);
\draw[ thick, fill=myblue4, rounded corners=2pt] (-1.4,0.4) rectangle (-.9,-0.1);
\draw[thick] (-1.15,0) -- (-1,0) -- (-1,0.15);
\draw[thick] (-2.25,0.5) -- (-1.25,-0.5);
\draw[thick] (-2.25,-0.5) -- (-1.25,0.5);
\draw[ thick, fill=myorange, rounded corners=2pt] (-2,0.25) rectangle (-1.5,-0.25);
\draw[thick] (-1.75,0.15) -- (-1.6,0.15) -- (-1.6,0);
\Text[x=-2.25,y=-0.6]{}
\end{tikzpicture}
= W\otimes W^{*}\,, \\
\nonumber &
\begin{tikzpicture}[baseline=(current  bounding  box.center), scale=.7]
\draw[very thick] (-4.25,0.5) -- (-4.25,-0.5);
\draw[ thick, fill=myYO, rounded corners=2pt] (-4.25,0) circle (.15);
\draw[thick, rotate around = {-45:(0.525-4.77,0.375-0.4)}]  (.45-4.77,0.3-0.4) -- (.45-4.77,0.45-0.4) -- (.6-4.77,0.45-0.4);
\Text[x=-4.25,y=-0.75]{}
\end{tikzpicture}
=
\begin{tikzpicture}[baseline=(current  bounding  box.center), scale=.7]
\draw[ thick] (-4,0.5) -- (-4,-0.5);
\draw[ thick, fill=mygray4, rounded corners=2pt] (-4,0) circle (.15);
\draw[thick, rotate around = {135:(0.525-4.27-0.25,0.375-0.35)}]  (.45-4.27-.25,0.3-0.35) -- (.45-4.27-.25,0.45-0.35) -- (.6-4.27-.25,0.45-0.35);
\draw[ thick] (-4.25,0.5) -- (-4.25,-0.5);
\draw[ thick, fill=myblue10, rounded corners=2pt] (-4.25,0) circle (.15);
\draw[thick, rotate around = {-45:(0.525-4.77,0.375-0.4)}]  (.45-4.77,0.3-0.4) -- (.45-4.77,0.45-0.4) -- (.6-4.77,0.45-0.4);
\Text[x=-4.25,y=-0.75]{}
\end{tikzpicture}
= u_{x}\otimes u_{x}^{*},\,\, w_{x}\otimes w_{x}^{*}\,.
\end{eqnarray}
Here and in the following $(\cdot)^*$ denotes complex conjugation in the canonical basis \eqref{eq:realbasis}.

\subsection{Local disorder averaging} 
\label{sec:Ave}

As mentioned in Sec.~\ref{sec:SFF} the definition of SFF requires an average. Since our interest is mainly on clean systems, we consider averages over \emph{onsite disorder} that can be made arbitrary weak. This kind of disorder is arguably the most harmless form of disorder that one can introduce in the system because it does not couple different sites. In particular, we focus on the following generic model of on-site disorder where the local gates \eqref{eq:Floquetgates1}--\eqref{eq:Floquetgates2} are specified by fixed unitary interactions $U, W \in {\rm U}(d^2)$, and site-dependent local gates $u_{x}, w_{x} \in {\rm SU}(d)$ of the general form
\be
u_{x} = e^{i \boldsymbol{\theta}_{0,x}\cdot\boldsymbol{\sigma}},
\qquad
w_{x} = e^{i \boldsymbol{\theta}_{1,x}\cdot\boldsymbol{\sigma}^T},
\qquad 
x\in\Lambda_L,\;\;\boldsymbol{\theta}_{\iota,x}\in\mathbb R^{d^2-1}\,.
\label{eq:urx}
\ee
The vector $\boldsymbol{\sigma}=(\sigma_1,\sigma_2,\ldots, \sigma_{d^2-1})$ is formed by Generalised Gell-Mann matrices $\sigma_a$~\cite{GenGellMann} (Pauli matrices for $d=2$, Gell-Mann matrices for $d=3$, etc.), the Hermitian generators of $\mathfrak{su}(d)$, and $\boldsymbol{\sigma}^T=(\sigma_1^T,\sigma_2^T,\ldots, \sigma_{d^2-1}^T)$ is the vector of the corresponding transposed generators. The expectation can be explicitly written in terms of a factorised measure as:
\be
\mathbb E[f] = \int  f(\boldsymbol\theta)  \prod_{x=0}^{L-1}\prod_{\iota,\iota'=0}^1 
g_{\iota\iota'}(\boldsymbol{\theta}_{\iota,x+\frac{\iota'}{2}}) {\rm d}^{d^2-1}\boldsymbol{\theta}_{\iota, x+\frac{\iota'}{2}}
\,,
\quad
\boldsymbol \theta \equiv (\boldsymbol{\theta}_{\iota,x})^{\iota=0,1}_{x\in\Lambda_L}\,.
\label{eq:measure}
\ee 
where $g_{\iota\iota'}\in L^1[\mathbb R^{d^2-1}]$ are arbitrary probability densities of i.i.d. random variables $\boldsymbol{\theta}_{\iota,x}$. Note that distributions on integer ($\iota'=0$) and half-odd-integer ($\iota'=1$) sublattices are generally different.

\subsection{Space-time duality}
\label{sec:duality}

The key property of $\mathbb E[\cdot]$ (\ref{eq:measure}) is the factorization with respect to a spatial coordinate $x$. This means that, even though the diagram \eqref{eq:SFFfolded} cannot be thought of as the trace of the product of $t$ transfer matrices in the time direction (because the average couples different time layers), 
it can be thought of as the trace of the product of $L$ transfer matrices in the space direction. Specifically, 
\be
K(t,L)=\begin{tikzpicture}[baseline=(current  bounding  box.center), scale=0.5]
\Text[x=-10,y=4.15]{}
\foreach \i in{1.5,2.5,3.5}{
\draw[very thick] (-10+0.5+0,2*\i-0.5-3.5) arc (45:270:0.15);
}
\foreach \i in{0.5,1.5,2.5}
{
\draw[very thick] (-10+0.5,1+2*\i-0.5-3.5) arc (315:90:0.15);
}
\end{tikzpicture}\,\mathbb{E}\Bigl[\begin{tikzpicture}[baseline=(current  bounding  box.center), scale=0.5]
\foreach \i in {5}{
\draw[very thick, dotted] (2*\i+2-12.5+0.255,-1.75-0.1) -- (2*\i+2-12.5+0.255,4.25-0.1);
\draw[very thick, dotted] (2*\i+2-11.5-0.255,-1.75-0.1) -- (2*\i+2-11.5-0.255,4.25-0.1);}

\foreach \i in {5}{
\draw[very thick] (2*\i+2-11.5,4) arc (-45:175:0.15);
\draw[very thick] (2*\i+2-11.5,-2) arc (315:180:0.15);
\draw[very thick] (2*\i+2-0.5-12,-2) arc (-135:0:0.15);
}
\foreach \i in {6}{
\draw[very thick] (2*\i+2-2.5-12,4) arc (225:0:0.15);
}
\foreach \jj[evaluate=\jj as \j using -2*(ceil(\jj/2)-\jj/2)] in {-1,-3,-5}{
\foreach \i in {1}
{
\draw[very thick] (.5-2*\i-1*\j,-2-1*\jj) -- (1-2*\i-1*\j,-1.5-\jj);
\draw[very thick] (1-2*\i-1*\j,-1.5-1*\jj) -- (1.5-2*\i-1*\j,-2-\jj);
\draw[very thick] (.5-2*\i-1*\j,-1-1*\jj) -- (1-2*\i-1*\j,-1.5-\jj);
\draw[very thick] (1-2*\i-1*\j,-1.5-1*\jj) -- (1.5-2*\i-1*\j,-1-\jj);
\draw[thick, fill=OliveGreen, rounded corners=2pt] (0.75-2*\i-1*\j,-1.75-\jj) rectangle (1.25-2*\i-1*\j,-1.25-\jj);
\draw[ thick] (-2*\i+2,-1.35-\jj) -- (-2*\i+2.15,-1.35-\jj) -- (-2*\i+2.15,-1.5-\jj);%
}
}
\foreach \jj[evaluate=\jj as \j using -2*(ceil(\jj/2)-\jj/2)] in {-4,-2,0}{
\foreach \i in {1}
{
\draw[very thick] (.5-2*\i-1*\j,-2-1*\jj) -- (1-2*\i-1*\j,-1.5-\jj);
\draw[very thick] (1-2*\i-1*\j,-1.5-1*\jj) -- (1.5-2*\i-1*\j,-2-\jj);
\draw[very thick] (.5-2*\i-1*\j,-1-1*\jj) -- (1-2*\i-1*\j,-1.5-\jj);
\draw[very thick] (1-2*\i-1*\j,-1.5-1*\jj) -- (1.5-2*\i-1*\j,-1-\jj);
\draw[thick, fill=mygreen, rounded corners=2pt] (0.75-2*\i-1*\j,-1.75-\jj) rectangle (1.25-2*\i-1*\j,-1.25-\jj);
\draw[ thick] (-2*\i+1,-1.35-\jj) -- (-2*\i+1.15,-1.35-\jj) -- (-2*\i+1.15,-1.5-\jj);%
}
}
\foreach \jj in {0,2,4}{
\draw[ thick, fill=myyellow1, rounded corners=2pt] (0.5,-1+\jj) circle (.15);
\draw[ thick, fill=myyellow2, rounded corners=2pt] (-0.5,-1+\jj) circle (.15);
\draw[ thick, fill=myyellow8, rounded corners=2pt] (0.5,\jj) circle (.15);
\draw[ thick, fill=myyellow9, rounded corners=2pt] (-0.5,\jj) circle (.15);
}
\foreach \jj in {0,-2,-4}{
\foreach \i in {1}{
\draw[ thick] (-2*\i+1.4,-.95-\jj) -- (-2*\i+1.55,-.95-\jj) -- (-2*\i+1.55,-1.1-\jj);}
\foreach \i in {1}{
\draw[ thick] (-2*\i+2.4,0.05-\jj) -- (-2*\i+2.55,0.05-\jj) -- (-2*\i+2.55,-0.1-\jj);}
\foreach \i in {1}{
\draw[ thick]  (-2*\i+1.45,-0.1-\jj) -- (-2*\i+1.45,0.05-\jj) -- (-2*\i+1.6,0.05-\jj);}
\foreach \i in {0}{
\draw[ thick]  (-2*\i+.45,-1.1-\jj) -- (-2*\i+.45,-0.95-\jj) -- (-2*\i+.6,-0.95-\jj);}
}
\end{tikzpicture}\Bigr]\mathbb{E}\Bigl[\begin{tikzpicture}[baseline=(current  bounding  box.center), scale=0.5]
\foreach \i in {5}{
\draw[very thick, dotted] (2*\i+2-12.5+0.255,-1.75-0.1) -- (2*\i+2-12.5+0.255,4.25-0.1);
\draw[very thick, dotted] (2*\i+2-11.5-0.255,-1.75-0.1) -- (2*\i+2-11.5-0.255,4.25-0.1);}

\foreach \i in {5}{
\draw[very thick] (2*\i+2-11.5,4) arc (-45:175:0.15);
\draw[very thick] (2*\i+2-11.5,-2) arc (315:180:0.15);
\draw[very thick] (2*\i+2-0.5-12,-2) arc (-135:0:0.15);
}
\foreach \i in {6}{
\draw[very thick] (2*\i+2-2.5-12,4) arc (225:0:0.15);
}
\foreach \jj[evaluate=\jj as \j using -2*(ceil(\jj/2)-\jj/2)] in {-1,-3,-5}{
\foreach \i in {1}
{
\draw[very thick] (.5-2*\i-1*\j,-2-1*\jj) -- (1-2*\i-1*\j,-1.5-\jj);
\draw[very thick] (1-2*\i-1*\j,-1.5-1*\jj) -- (1.5-2*\i-1*\j,-2-\jj);
\draw[very thick] (.5-2*\i-1*\j,-1-1*\jj) -- (1-2*\i-1*\j,-1.5-\jj);
\draw[very thick] (1-2*\i-1*\j,-1.5-1*\jj) -- (1.5-2*\i-1*\j,-1-\jj);
\draw[thick, fill=OliveGreen, rounded corners=2pt] (0.75-2*\i-1*\j,-1.75-\jj) rectangle (1.25-2*\i-1*\j,-1.25-\jj);
\draw[ thick] (-2*\i+2,-1.35-\jj) -- (-2*\i+2.15,-1.35-\jj) -- (-2*\i+2.15,-1.5-\jj);%
}
}
\foreach \jj[evaluate=\jj as \j using -2*(ceil(\jj/2)-\jj/2)] in {-4,-2,0}{
\foreach \i in {1}
{
\draw[very thick] (.5-2*\i-1*\j,-2-1*\jj) -- (1-2*\i-1*\j,-1.5-\jj);
\draw[very thick] (1-2*\i-1*\j,-1.5-1*\jj) -- (1.5-2*\i-1*\j,-2-\jj);
\draw[very thick] (.5-2*\i-1*\j,-1-1*\jj) -- (1-2*\i-1*\j,-1.5-\jj);
\draw[very thick] (1-2*\i-1*\j,-1.5-1*\jj) -- (1.5-2*\i-1*\j,-1-\jj);
\draw[thick, fill=mygreen, rounded corners=2pt] (0.75-2*\i-1*\j,-1.75-\jj) rectangle (1.25-2*\i-1*\j,-1.25-\jj);
\draw[ thick] (-2*\i+1,-1.35-\jj) -- (-2*\i+1.15,-1.35-\jj) -- (-2*\i+1.15,-1.5-\jj);%
}
}
\foreach \jj in {0,2,4}{
\draw[ thick, fill=myyellow1, rounded corners=2pt] (0.5,-1+\jj) circle (.15);
\draw[ thick, fill=myyellow2, rounded corners=2pt] (-0.5,-1+\jj) circle (.15);
\draw[ thick, fill=myyellow8, rounded corners=2pt] (0.5,\jj) circle (.15);
\draw[ thick, fill=myyellow9, rounded corners=2pt] (-0.5,\jj) circle (.15);
}
\foreach \jj in {0,-2,-4}{
\foreach \i in {1}{
\draw[ thick] (-2*\i+1.4,-.95-\jj) -- (-2*\i+1.55,-.95-\jj) -- (-2*\i+1.55,-1.1-\jj);}
\foreach \i in {1}{
\draw[ thick] (-2*\i+2.4,0.05-\jj) -- (-2*\i+2.55,0.05-\jj) -- (-2*\i+2.55,-0.1-\jj);}
\foreach \i in {1}{
\draw[ thick]  (-2*\i+1.45,-0.1-\jj) -- (-2*\i+1.45,0.05-\jj) -- (-2*\i+1.6,0.05-\jj);}
\foreach \i in {0}{
\draw[ thick]  (-2*\i+.45,-1.1-\jj) -- (-2*\i+.45,-0.95-\jj) -- (-2*\i+.6,-0.95-\jj);}
}
\end{tikzpicture}\Bigr]\mathbb{E}\Bigl[\begin{tikzpicture}[baseline=(current  bounding  box.center), scale=0.5]
\foreach \i in {5}{
\draw[very thick, dotted] (2*\i+2-12.5+0.255,-1.75-0.1) -- (2*\i+2-12.5+0.255,4.25-0.1);
\draw[very thick, dotted] (2*\i+2-11.5-0.255,-1.75-0.1) -- (2*\i+2-11.5-0.255,4.25-0.1);}

\foreach \i in {5}{
\draw[very thick] (2*\i+2-11.5,4) arc (-45:175:0.15);
\draw[very thick] (2*\i+2-11.5,-2) arc (315:180:0.15);
\draw[very thick] (2*\i+2-0.5-12,-2) arc (-135:0:0.15);
}
\foreach \i in {6}{
\draw[very thick] (2*\i+2-2.5-12,4) arc (225:0:0.15);
}
\foreach \jj[evaluate=\jj as \j using -2*(ceil(\jj/2)-\jj/2)] in {-1,-3,-5}{
\foreach \i in {1}
{
\draw[very thick] (.5-2*\i-1*\j,-2-1*\jj) -- (1-2*\i-1*\j,-1.5-\jj);
\draw[very thick] (1-2*\i-1*\j,-1.5-1*\jj) -- (1.5-2*\i-1*\j,-2-\jj);
\draw[very thick] (.5-2*\i-1*\j,-1-1*\jj) -- (1-2*\i-1*\j,-1.5-\jj);
\draw[very thick] (1-2*\i-1*\j,-1.5-1*\jj) -- (1.5-2*\i-1*\j,-1-\jj);
\draw[thick, fill=OliveGreen, rounded corners=2pt] (0.75-2*\i-1*\j,-1.75-\jj) rectangle (1.25-2*\i-1*\j,-1.25-\jj);
\draw[ thick] (-2*\i+2,-1.35-\jj) -- (-2*\i+2.15,-1.35-\jj) -- (-2*\i+2.15,-1.5-\jj);%
}
}
\foreach \jj[evaluate=\jj as \j using -2*(ceil(\jj/2)-\jj/2)] in {-4,-2,0}{
\foreach \i in {1}
{
\draw[very thick] (.5-2*\i-1*\j,-2-1*\jj) -- (1-2*\i-1*\j,-1.5-\jj);
\draw[very thick] (1-2*\i-1*\j,-1.5-1*\jj) -- (1.5-2*\i-1*\j,-2-\jj);
\draw[very thick] (.5-2*\i-1*\j,-1-1*\jj) -- (1-2*\i-1*\j,-1.5-\jj);
\draw[very thick] (1-2*\i-1*\j,-1.5-1*\jj) -- (1.5-2*\i-1*\j,-1-\jj);
\draw[thick, fill=mygreen, rounded corners=2pt] (0.75-2*\i-1*\j,-1.75-\jj) rectangle (1.25-2*\i-1*\j,-1.25-\jj);
\draw[ thick] (-2*\i+1,-1.35-\jj) -- (-2*\i+1.15,-1.35-\jj) -- (-2*\i+1.15,-1.5-\jj);%
}
}
\foreach \jj in {0,2,4}{
\draw[ thick, fill=myyellow1, rounded corners=2pt] (0.5,-1+\jj) circle (.15);
\draw[ thick, fill=myyellow2, rounded corners=2pt] (-0.5,-1+\jj) circle (.15);
\draw[ thick, fill=myyellow8, rounded corners=2pt] (0.5,\jj) circle (.15);
\draw[ thick, fill=myyellow9, rounded corners=2pt] (-0.5,\jj) circle (.15);
}
\foreach \jj in {0,-2,-4}{
\foreach \i in {1}{
\draw[ thick] (-2*\i+1.4,-.95-\jj) -- (-2*\i+1.55,-.95-\jj) -- (-2*\i+1.55,-1.1-\jj);}
\foreach \i in {1}{
\draw[ thick] (-2*\i+2.4,0.05-\jj) -- (-2*\i+2.55,0.05-\jj) -- (-2*\i+2.55,-0.1-\jj);}
\foreach \i in {1}{
\draw[ thick]  (-2*\i+1.45,-0.1-\jj) -- (-2*\i+1.45,0.05-\jj) -- (-2*\i+1.6,0.05-\jj);}
\foreach \i in {0}{
\draw[ thick]  (-2*\i+.45,-1.1-\jj) -- (-2*\i+.45,-0.95-\jj) -- (-2*\i+.6,-0.95-\jj);}
}
\end{tikzpicture}\Bigr]\mathbb{E}\Bigl[\begin{tikzpicture}[baseline=(current  bounding  box.center), scale=0.5]
\foreach \i in {5}{
\draw[very thick, dotted] (2*\i+2-12.5+0.255,-1.75-0.1) -- (2*\i+2-12.5+0.255,4.25-0.1);
\draw[very thick, dotted] (2*\i+2-11.5-0.255,-1.75-0.1) -- (2*\i+2-11.5-0.255,4.25-0.1);}

\foreach \i in {5}{
\draw[very thick] (2*\i+2-11.5,4) arc (-45:175:0.15);
\draw[very thick] (2*\i+2-11.5,-2) arc (315:180:0.15);
\draw[very thick] (2*\i+2-0.5-12,-2) arc (-135:0:0.15);
}
\foreach \i in {6}{
\draw[very thick] (2*\i+2-2.5-12,4) arc (225:0:0.15);
}
\foreach \jj[evaluate=\jj as \j using -2*(ceil(\jj/2)-\jj/2)] in {-1,-3,-5}{
\foreach \i in {1}
{
\draw[very thick] (.5-2*\i-1*\j,-2-1*\jj) -- (1-2*\i-1*\j,-1.5-\jj);
\draw[very thick] (1-2*\i-1*\j,-1.5-1*\jj) -- (1.5-2*\i-1*\j,-2-\jj);
\draw[very thick] (.5-2*\i-1*\j,-1-1*\jj) -- (1-2*\i-1*\j,-1.5-\jj);
\draw[very thick] (1-2*\i-1*\j,-1.5-1*\jj) -- (1.5-2*\i-1*\j,-1-\jj);
\draw[thick, fill=OliveGreen, rounded corners=2pt] (0.75-2*\i-1*\j,-1.75-\jj) rectangle (1.25-2*\i-1*\j,-1.25-\jj);
\draw[ thick] (-2*\i+2,-1.35-\jj) -- (-2*\i+2.15,-1.35-\jj) -- (-2*\i+2.15,-1.5-\jj);%
}
}
\foreach \jj[evaluate=\jj as \j using -2*(ceil(\jj/2)-\jj/2)] in {-4,-2,0}{
\foreach \i in {1}
{
\draw[very thick] (.5-2*\i-1*\j,-2-1*\jj) -- (1-2*\i-1*\j,-1.5-\jj);
\draw[very thick] (1-2*\i-1*\j,-1.5-1*\jj) -- (1.5-2*\i-1*\j,-2-\jj);
\draw[very thick] (.5-2*\i-1*\j,-1-1*\jj) -- (1-2*\i-1*\j,-1.5-\jj);
\draw[very thick] (1-2*\i-1*\j,-1.5-1*\jj) -- (1.5-2*\i-1*\j,-1-\jj);
\draw[thick, fill=mygreen, rounded corners=2pt] (0.75-2*\i-1*\j,-1.75-\jj) rectangle (1.25-2*\i-1*\j,-1.25-\jj);
\draw[ thick] (-2*\i+1,-1.35-\jj) -- (-2*\i+1.15,-1.35-\jj) -- (-2*\i+1.15,-1.5-\jj);%
}
}
\foreach \jj in {0,2,4}{
\draw[ thick, fill=myyellow1, rounded corners=2pt] (0.5,-1+\jj) circle (.15);
\draw[ thick, fill=myyellow2, rounded corners=2pt] (-0.5,-1+\jj) circle (.15);
\draw[ thick, fill=myyellow8, rounded corners=2pt] (0.5,\jj) circle (.15);
\draw[ thick, fill=myyellow9, rounded corners=2pt] (-0.5,\jj) circle (.15);
}
\foreach \jj in {0,-2,-4}{
\foreach \i in {1}{
\draw[ thick] (-2*\i+1.4,-.95-\jj) -- (-2*\i+1.55,-.95-\jj) -- (-2*\i+1.55,-1.1-\jj);}
\foreach \i in {1}{
\draw[ thick] (-2*\i+2.4,0.05-\jj) -- (-2*\i+2.55,0.05-\jj) -- (-2*\i+2.55,-0.1-\jj);}
\foreach \i in {1}{
\draw[ thick]  (-2*\i+1.45,-0.1-\jj) -- (-2*\i+1.45,0.05-\jj) -- (-2*\i+1.6,0.05-\jj);}
\foreach \i in {0}{
\draw[ thick]  (-2*\i+.45,-1.1-\jj) -- (-2*\i+.45,-0.95-\jj) -- (-2*\i+.6,-0.95-\jj);}
}
\end{tikzpicture}\Bigr]\mathbb{E}\Bigl[\begin{tikzpicture}[baseline=(current  bounding  box.center), scale=0.5]
\foreach \i in {5}{
\draw[very thick, dotted] (2*\i+2-12.5+0.255,-1.75-0.1) -- (2*\i+2-12.5+0.255,4.25-0.1);
\draw[very thick, dotted] (2*\i+2-11.5-0.255,-1.75-0.1) -- (2*\i+2-11.5-0.255,4.25-0.1);}

\foreach \i in {5}{
\draw[very thick] (2*\i+2-11.5,4) arc (-45:175:0.15);
\draw[very thick] (2*\i+2-11.5,-2) arc (315:180:0.15);
\draw[very thick] (2*\i+2-0.5-12,-2) arc (-135:0:0.15);
}
\foreach \i in {6}{
\draw[very thick] (2*\i+2-2.5-12,4) arc (225:0:0.15);
}
\foreach \jj[evaluate=\jj as \j using -2*(ceil(\jj/2)-\jj/2)] in {-1,-3,-5}{
\foreach \i in {1}
{
\draw[very thick] (.5-2*\i-1*\j,-2-1*\jj) -- (1-2*\i-1*\j,-1.5-\jj);
\draw[very thick] (1-2*\i-1*\j,-1.5-1*\jj) -- (1.5-2*\i-1*\j,-2-\jj);
\draw[very thick] (.5-2*\i-1*\j,-1-1*\jj) -- (1-2*\i-1*\j,-1.5-\jj);
\draw[very thick] (1-2*\i-1*\j,-1.5-1*\jj) -- (1.5-2*\i-1*\j,-1-\jj);
\draw[thick, fill=OliveGreen, rounded corners=2pt] (0.75-2*\i-1*\j,-1.75-\jj) rectangle (1.25-2*\i-1*\j,-1.25-\jj);
\draw[ thick] (-2*\i+2,-1.35-\jj) -- (-2*\i+2.15,-1.35-\jj) -- (-2*\i+2.15,-1.5-\jj);%
}
}
\foreach \jj[evaluate=\jj as \j using -2*(ceil(\jj/2)-\jj/2)] in {-4,-2,0}{
\foreach \i in {1}
{
\draw[very thick] (.5-2*\i-1*\j,-2-1*\jj) -- (1-2*\i-1*\j,-1.5-\jj);
\draw[very thick] (1-2*\i-1*\j,-1.5-1*\jj) -- (1.5-2*\i-1*\j,-2-\jj);
\draw[very thick] (.5-2*\i-1*\j,-1-1*\jj) -- (1-2*\i-1*\j,-1.5-\jj);
\draw[very thick] (1-2*\i-1*\j,-1.5-1*\jj) -- (1.5-2*\i-1*\j,-1-\jj);
\draw[thick, fill=mygreen, rounded corners=2pt] (0.75-2*\i-1*\j,-1.75-\jj) rectangle (1.25-2*\i-1*\j,-1.25-\jj);
\draw[ thick] (-2*\i+1,-1.35-\jj) -- (-2*\i+1.15,-1.35-\jj) -- (-2*\i+1.15,-1.5-\jj);%
}
}
\foreach \jj in {0,2,4}{
\draw[ thick, fill=myyellow1, rounded corners=2pt] (0.5,-1+\jj) circle (.15);
\draw[ thick, fill=myyellow2, rounded corners=2pt] (-0.5,-1+\jj) circle (.15);
\draw[ thick, fill=myyellow8, rounded corners=2pt] (0.5,\jj) circle (.15);
\draw[ thick, fill=myyellow9, rounded corners=2pt] (-0.5,\jj) circle (.15);
}
\foreach \jj in {0,-2,-4}{
\foreach \i in {1}{
\draw[ thick] (-2*\i+1.4,-.95-\jj) -- (-2*\i+1.55,-.95-\jj) -- (-2*\i+1.55,-1.1-\jj);}
\foreach \i in {1}{
\draw[ thick] (-2*\i+2.4,0.05-\jj) -- (-2*\i+2.55,0.05-\jj) -- (-2*\i+2.55,-0.1-\jj);}
\foreach \i in {1}{
\draw[ thick]  (-2*\i+1.45,-0.1-\jj) -- (-2*\i+1.45,0.05-\jj) -- (-2*\i+1.6,0.05-\jj);}
\foreach \i in {0}{
\draw[ thick]  (-2*\i+.45,-1.1-\jj) -- (-2*\i+.45,-0.95-\jj) -- (-2*\i+.6,-0.95-\jj);}
}
\end{tikzpicture}\Bigr]
\begin{tikzpicture}[baseline=(current  bounding  box.center), scale=0.5]
\Text[x=0.5,y=4.15]{}
\foreach \i in{1.5,2.5,3.5}{
\draw[very thick] (0.5,2*\i-0.5-3.5) arc (45:-90:0.15);
}
\foreach \i in{0.5,1.5,2.5}
{
\draw[very thick] (0.5,1+2*\i-0.5-3.5) arc (-45:90:0.15);
}
\end{tikzpicture}\,.
\label{eq:SFFTMspace}
\ee
In equations this is expressed as 
\bea
K(t,L) &=&  
 \mathbb E\left[\left({\rm tr}\,\mathbb U_{L}^t\right) \left({\rm tr}\,\mathbb U_{L}^t\right)^*\right] = 
 \mathbb E\left[ {\rm tr}\left(\mathbb U^{\phantom{*}}_{L}\otimes \mathbb U^*_{L}\right)^t\right] \nonumber\\
 &=& \mathbb E\left[ {\rm tr}\left(\prod_{x=1}^L \tilde{\mathbb U}_{t}^{\phantom{*}}(x) \otimes \tilde{\mathbb U}_{t}(x)^*\right)\right] =
 {\rm tr}\left(\mathbb E\left[\tilde{\mathbb U}_{t}^{\phantom{*}} \otimes \tilde{\mathbb U}_{t}^*\right]\right)^L \nonumber\\
 &=& {\rm tr}\,\mathbb T^L,
 \label{eq:SFFduality}
\eea
where the tensor product operates between the two different time sheets, and we introduced the following definitions:
\begin{itemize}
\item[(i)] ``Dual'' Floquet operator propagating in the space-direction over the Hilbert space ${\cal H}_{2t}$ of $2t$ qudits, explicitly depending on the position $x\in \mathbb Z_L$:
\be
\tilde{\mathbb U}_{t}(x) := \prod_{\tau \in \mathbb Z_{t}+\frac{1}{2}}\!\!\eta_{\tau,t}(\tilde U\,( u_{x-\frac{1}{2}} \otimes w_{x-\frac{1}{2}}^T)) \prod_{\tau \in \mathbb Z_{t}} \eta_{\tau,t}(\tilde W \, (w_{x} \otimes u_x^T))\,.
\ee
Here $\tilde{U}$, $\tilde{W} \in {\rm End}(\mathcal H_2)$ are the ``dual" 2-body interaction gates defined via the
space-time duality mapping $\,\tilde{} : {\rm End}(\mathcal H_2) \to  {\rm End}(\mathcal H_2)$.
Specifically, for any $O\in{\rm End}(\mathcal H_2)$
with matrix elements
\be
O_{i_1i_2,j_1j_2} = \bra{i_1}\otimes \bra{i_2} O \ket{j_1}\otimes \ket{j_2}\,,
\ee
we define
\be
\tilde{O}_{jl,ik} := O_{ij,kl},\quad
i,j,k,l\in\{0,1, \ldots, d-1\}\,.
\label{eq:duallocalgate}
\ee
We see that $\tilde{U}_{ij,kl}$ and $\tilde{W}_{ij,kl}$ correspond to a particular reshuffling of the indices of ${U}_{ij,kl}$ and ${W}_{ij,kl}$.  
\item[(ii)] SFF--transfer matrix:
\be
\mathbb T:= \mathbb E\bigl[\tilde{\mathbb U}_t^{\phantom{*}} \otimes \tilde{\mathbb U}_t^*\bigr]\in {\rm End}({\cal H}_{2t}\otimes{\cal H}_{2t}).
\label{eq:SFFTM}
\ee
  Note that $\mathbb T$ \emph{does not} depend on position $x$ due to the identical distribution of $(\boldsymbol{\theta}_{\iota,x-\frac{1}{2}},\boldsymbol{\theta}_{\iota,x})$ for all $x\in\mathbb Z_L$.
\end{itemize} 
More specifically, performing explicitly the average via \eqref{eq:measure}, we find 
\be
\mathbb T = (\tilde{\mathbb U} \otimes \tilde{\mathbb U}^*) \mathbb O^\dag_1 (\tilde{\mathbb W} \otimes \tilde{\mathbb W}^*) \mathbb O^{\phantom{\dag}}_0,
\label{eq:SFFTM2}
\ee
where we introduced  
\begin{align}
\tilde{\mathbb U} &:= \!\!\prod_{\tau \in \mathbb Z_{t}+\frac{1}{2}}\!\!\eta_{\tau,t}(\tilde U)\,, 
\label{eq:Utildee}\\
\tilde{\mathbb W} &:= \prod_{\tau \in \mathbb Z_{t}} \eta_{\tau,t}(\tilde W)\,,
\label{eq:Utildeo}\\
\mathbb O_{\iota'} &:= \mathbb O_{0\iota'} \mathbb O_{1\iota'}=  \mathbb O_{1\iota'} \mathbb O_{0\iota'}\,,\\  
\mathbb O_{\iota\iota'} &:= \int\!{\rm d}^{d^2-1}\boldsymbol{\theta}\,g_{\iota\iota'}(\boldsymbol{\theta})
\exp\left(i\boldsymbol\theta\cdot (\boldsymbol{M}_{\iota}\otimes \1_{2t}-\1_{2t}\otimes \boldsymbol{M}^*_{\iota})\right)\,.
\label{eq:defOa}
\end{align}
Here $\boldsymbol{M}_\iota = (M_{1,\iota},M_{2,\iota},\ldots, M_{d^2-1,\iota})$ with $M_{a, \iota}$ denoting the representation of the Hermitian generators of ${\mathfrak su}(d)$ in the lattice made of integer and half-odd-integer time indices 
\be
M_{a, \iota} :=\sum_{\tau \in \mathbb Z_t + \frac{1}{2}\iota} \!\!\!\!\!\!\sigma_{a,\tau}\,,\quad \sigma_{a,\tau} : =
\Pi^{2\tau}_{2t}(\sigma_a\otimes \1_{2t-1})\Pi^{-2\tau}_{2t}\,,\quad \iota\in\{0,1\}\,.
\label{eq:sublatticeM}
\ee
Normalisability and non-negativity of the probability densities $g_{\iota\iota'}$ imply the following important properties of the operators $\mathbb O_{\iota\iota'}$:
\begin{enumerate}
\item[(a)] $\mathbb O_{\iota\iota'}$ is a \emph{non-expansive mapping}:
\be
\!\!\!\| \mathbb O_{\iota\iota'} \| \le 
\int\!{\rm d}^{d^2-1}\boldsymbol{\theta}\, |g_{\iota\iota'}(\boldsymbol{\theta})|
\left\|\exp\left(i\boldsymbol\theta\cdot (\boldsymbol{M}_{\iota}\otimes \1_{2t}-\1_{2t}\otimes \boldsymbol{M}^*_{\iota})\right)\right\| = 1\,.
\label{eq:propi}
\ee
\item[(b)]
Let $\mathbb O_{\iota\iota'} \ket{B} = e^{i\phi} \ket{B}$ for some
$\ket{B}\in \mathcal H_{2t}\otimes \mathcal H_{2t}$ and $\phi\in\mathbb R$. Then:
\be
\phi = 0,\,\, \textrm{and}\,\,
(M_{a,\iota}\otimes \1_{2t}-\1_{2t}\otimes M^*_{a,\iota})\ket{B} = 0\,,\,\, \forall a\in\{1,2,\ldots,d^2-1\}\,.
\label{eq:propii}
\ee
Indeed, the assumptions imply 
\be
e^{i\boldsymbol\theta\cdot (\boldsymbol{M}_{\iota}\otimes \1_{2t}-\1_{2t}\otimes \boldsymbol{M}^*_{\iota})}\ket{B} = e^{i\phi} \ket{B},
\label{eq:propsup}
\ee
for a dense set $\mathcal S \ni \boldsymbol{\theta}$ with positive measure, e.g. the support of $g_{\iota\iota'}$ or its dense subset. Without loss of generality we can assume that $\mathcal S$ contains the origin $\boldsymbol 0\in\mathcal S$.
Taking a partial derivative $\frac{\partial}{\partial\theta_a}|_{\boldsymbol\theta=\boldsymbol 0}$ of (\ref{eq:propsup}) we obtain (\ref{eq:propii}).
\end{enumerate}

As convenient examples we can consider a Gaussian measure 
\be
g_{\iota\iota'}(\boldsymbol \theta)=
\prod_{a=1}^{d^2-1}\frac{1}{\sqrt{2\pi}\nu_{a\iota\iota'}}
\exp\left(-{\frac{1}{2}} \frac{\theta_a^2}{\nu^2_{a\iota\iota'}}\right)\,,
\ee
or a box-measure
\be
g_{\iota\iota'}(\boldsymbol \theta)=
\prod_{a=1}^{d^2-1} \frac{1}{2\nu_{a\iota\iota'}}
\Theta(\nu_{a\iota\iota'}-|\theta_a|)\,,
\ee
where choosing sufficiently small nonvanishing variabilities $\nu_{a\iota\iota'} > 0$ allows for arbitrary concentration of measure around the identity in ${\rm SU}(d)$, and hence description of an ``almost clean system''.

\subsection{Relevant limits}

For the class of dual-unitary circuits, which is the main focus of this paper and will be elaborated in the next section,  
the expression  (\ref{eq:SFFduality}) in terms of the map $\mathbb T$  [which is, in fact, a vectorised form of a completely positive trace preserving and unital mapping over ${\rm End}(\mathcal{H}_{2t})$] allows us to explicitly compute the SFF at any fixed $t$ in the thermodynamic limit ${L\to\infty}$. In particular, as we show below, we find that 
\be
\lim_{L\to\infty}K(t,L) = \lim_{\mathcal{ N}\to\infty}K_{\rm RMT}(t,\mathcal{N})=t,\qquad \forall t.
\ee

This fact is quite remarkable and signals a special property of the dual-unitary systems considered here. Indeed, in generic quantum chaotic systems one typically observes that there exists a timescale $t^*(L)$ such that the universal behaviour described by RMT emerges only for times ${t>t^*(L)}$. This timescale, usually referred to as the Thouless (or Ehrenfest) time, is typically observed to grow  monotonically with $L$~\cite{ljubotina,chalker,largeq}. The fact that for us, instead, $t^*(L)$ is strictly equal to zero can be interpreted as a sort of ``critical chaotic" (scale-free) property of dual-unitary systems.

It would certainly be desirable to address spectral correlations on other scales. For instance, those on the scale of mean level spacing (which becomes exponentially small in $L$ for a many-body system) and correspond to times $t$ of the order of Heisenberg time $\mathcal N = 2^L$. This would require studying the scaling limit 
\be
\mathcal{K}(\tau) = 
\lim_{L\to\infty} 2^{-L} K\left(\lfloor 2^L \tau\rfloor,L\right).
\ee
At the moment, however, we do not foresee any method to rigorously attack this challenging issue.

\section{Statement of the main results}
\label{sec:main}

\subsection{Exact SFF at large $L$ for dual-unitary circuits}
\label{sec:leadingeigs}

Using the representation \eqref{eq:SFFduality} we see that to compute the SFF one has to determine the spectrum of $\mathbb T$. In particular, to obtain $K(t,L)$ in the large size limit $L\to\infty$, it is sufficient to find all the eigenvalues with maximal magnitude.  

To achieve this goal we consider a special class of local quantum circuits called \emph{dual-unitary} circuits~\cite{BKP19}. These systems are characterised by the property that their dual local gates (cf. \eqref{eq:duallocalgate}) are unitary. Specifically, we consider local gates $U$ and $W$ (cf.~(\ref{eq:Floquetgates1}, \ref{eq:Floquetgates2})) that simultaneously fulfil 
\begin{align}
&U^\dag U = U U^{\dag}=\1, & &\tilde U^\dag \tilde U = \tilde U \tilde U^{\dag}=\1\,,\label{eq:dualunitarityU}\\
&W^\dag W= W W^{\dag}=\1, & &\tilde W^\dag \tilde W = \tilde W \tilde W^{\dag}=\1\,.
\label{eq:dualunitarityW}
\end{align}
The above conditions admit non-trivial solutions for any local dimension $d\geq2$~\cite{Gutkin20,ClLa20b}. A complete classification of solutions, however, has been achieved only for $d=2$~\cite{BKP19}.

An immediate consequence of \eqref{eq:dualunitarityU} and \eqref{eq:dualunitarityW} is that both $\tilde{\mathbb U}$ \eqref{eq:Utildee} and $\tilde{\mathbb W}$ \eqref{eq:Utildeo} are unitary.
This allows us to prove the following Lemma:
\begin{lemma}
\label{lemma1}
For dual-unitary circuits the matrix $\mathbb T$ (\ref{eq:SFFTM}) fulfils the properties:
\begin{itemize}
\item[(i)] $|\lambda|\leq 1$ for all $\lambda\in {\rm spect}(\mathbb T)$.
\item[(ii)] If $\mathbb T\!\ket{A} =e^{i \phi} \!\ket{A}$, $\phi\in\mathbb R$, then
\bea
&&(\tilde{\mathbb U} \otimes \tilde{\mathbb U}^*)\cdot (\tilde{\mathbb W} \otimes \tilde{\mathbb W}^*)\ket{A}=e^{i \phi}\!\ket{A}\,,\nonumber \\
&&
( M_{a, \iota}\otimes \1_{2t}-\1_{2t}\otimes  M_{a, \iota}^*)(\tilde{\mathbb W} \otimes \tilde{\mathbb W}^*) \ket{A} = 0\,,\label{eq:conditionsstateA}\\
&&
( M_{a, \iota}\otimes \1_{2t}-\1_{2t}\otimes  M_{a, \iota}^*)\ket{A} = 0, \quad \iota\in\{0,1\},\;a\in\{1,2,\ldots,d^2-1\}\,. \nonumber
\eea 
\item[(iii)] For any unimodular eigenvalue $\lambda$, $|\lambda| =1$, its algebraic and geometric multiplicities coincide (i.e. its Jordan blocks are trivial). 
\end{itemize}
\end{lemma} 
In essence, (i) and (iii) mean that $\mathbb T$ is a linear non-expansive mapping over $\mathcal H_{2t}\otimes \mathcal H_{2t}$, while (ii) suggests that computation of invariant subspaces can be reduced to simpler algebraic problems. Indeed, it ensures that for dual-unitary circuits 
\be
\mathcal N(\phi):=\frac{1}{L} \sum_{\ell = 1}^L e^{-i\phi \ell} K(t,\ell)
\ee 
approaches a finite value when $L\to\infty$ which is given by the number of linearly independent solutions $\ket{A}$ of the system of equations \eqref{eq:conditionsstateA}. The number of solutions for $\phi=0$ give the SFF averaged over the system size, while showing that $\phi=0$ is the only phase for which there are nontrivial solutions gives the thermodynamic limit $\lim_{L\to\infty} K(t,L)$.

In order to further simplify the conditions (\ref{eq:conditionsstateA}) we introduce a vector-operator isomorphism $\mathcal H \otimes \mathcal H\overset{vo}{\longleftrightarrow} {\rm End}(\mathcal H)$. Specifically, we define in the canonical basis \eqref{eq:realbasis}:
\be
\ket{j}\otimes \ket{j'} \overset{vo}{\longleftrightarrow} \ket{j}\bra{j'}\,.
\ee
This implies that the problem of finding all linearly independent states $\ket{A}$ solving \eqref{eq:conditionsstateA} is mapped to the one of finding all linearly 
independent operators $A$ satisfying, for all $a\in\{1,2,\ldots, d^2-1\}$ and $\iota \in\{0,1\}$:
\be
\tilde{\mathbb U} \tilde{\mathbb W} A \tilde{\mathbb W}^\dag \tilde{\mathbb U}^\dag =e^{i \phi} A,\quad [M_{a, \iota},A] = 0, \quad [\tilde{\mathbb W}^\dag M_{a, \iota} \tilde{\mathbb W},A]=0\,.  \label{eq:CondSU2}
\ee

The above conditions can be simplified further by making use of an explicit parametrisation of a dual-unitary matrices. Specifically, we parametrise the matrix $D \in  {\rm End}(\mathbb C^d\otimes \mathbb C^d)$ fulfilling $D  D^\dag= D^\dag  D=\1$ and $\tilde D \tilde D^\dag=\tilde D^\dag \tilde D=\1$ as follows  
\be
D = (u_1\otimes u_2) S e^{i J s_3 \otimes s_3} (u_3\otimes u_4), \qquad\qquad J\in[0,\pi],
\label{eq:parametrisationgen}
\ee
where $u_j\in {\rm U}(d)$ are arbitrary unitary matrices (`local gates') and $S\in {\rm End}(\mathbb C^d\otimes \mathbb C^d)$ is the SWAP operator defined as 
\be
S \ket{j_1}\otimes \ket{j_2} = \ket{j_2}\otimes \ket{j_1}\qquad \forall j_1,j_2\in\{0,1,\ldots d-1\}\,.
\ee
Finally, here and in the following $s_1,s_2,s_3$ designate the `spin matrices' carrying $d-$dimensional irreducible representation of $SU(2)$ over $\mathcal H_1$, satisfying
\be
[s_a,s_b]= i\sum_{c=1}^3 \epsilon_{abc} s_c,
\ee
where $\epsilon_{abc}$ is the three-dimensional Levi-Civita tensor,
and we choose $s_3$ to be diagonal in the canonical basis
\eqref{eq:realbasis}
\be
s_3 = {\rm diag}\left( -\frac{d-1}{2},-\frac{d-3}{2},\ldots,\frac{d-1}{2}\right)\,.
\label{eq:diag}
\ee
Local embeddings into ${\rm End}(\mathcal H_{2t})$ are, like in \eqref{eq:sublatticeM}, defined as
\be
s_{a,\tau} : = \Pi^{2\tau}_{2t}(s_a\otimes \1_{2t-1})\Pi^{-2\tau}_{2t}\,.
\ee
Note that for $d=2$ the parametrisation (\ref{eq:parametrisationgen}) exhausts all dual-unitary circuits~\cite{BKP19} while for $d>2$ it characterises a physically interesting sub-class~\cite{ClLa20b}. 

Plugging \eqref{eq:parametrisationgen} in the definitions  \eqref{eq:Utildee}--\eqref{eq:Utildeo} we find 
\begin{align}
&\tilde{\mathbb U} = e^{i \theta} 
e^{i \boldsymbol\alpha_0\cdot \boldsymbol M_0} 
e^{i \boldsymbol\alpha_1\cdot \boldsymbol M_1} \,\tilde{\mathbb V} \,  
e^{i \boldsymbol\beta_0\cdot \boldsymbol M_0} 
e^{i \boldsymbol\beta_1\cdot \boldsymbol M_1},\nonumber \\
&\tilde{\mathbb W} = 
e^{i \theta'} 
e^{i \boldsymbol\gamma_0\cdot \boldsymbol M_0} 
e^{i \boldsymbol\gamma_1\cdot \boldsymbol M_1} \,\tilde{\mathbb V}' \,  
e^{i \boldsymbol\delta_0\cdot \boldsymbol M_0} 
e^{i \boldsymbol\delta_1\cdot \boldsymbol M_1}, \label{eq:UGE}
\end{align}
where $\boldsymbol{\alpha}_\iota,\boldsymbol{\beta}_\iota,\boldsymbol{\gamma}_\iota,\boldsymbol{\delta}_\iota\in\mathbb R^{d^2-1}$, $\theta,\theta'\in\mathbb R$,
and we introduced 
\begin{align}
&\tilde{\mathbb V} :=  (S e^{i J s_3 \otimes s_3})^{\otimes t}\,, \label{eq:genmag}\\
&\tilde{\mathbb V}' :=  \Pi_{2t} (S e^{i J' s_3 \otimes s_3})^{\otimes t} \Pi^\dag_{2t}\,.\nonumber
\end{align}
We are then able to simplify the conditions for the existence of unimodular eigenvalues and write their invariant eigenoperator spaces in terms of a simple algebraic commutant:
\begin{lemma}
\label{lemma2}
For $J, J'\neq 0$ the conditions \eqref{eq:CondSU2} cannot be met unless $\phi=0$. In this case, they are equivalent to 
\be
[A, M_{a,\iota}]=0, \quad [A, M_{ab,\iota}]=0\,, \qquad a,b\in\{1,2,\ldots,d^2-1\}, \; \iota\in\{0, 1\}\,.
\label{eq:finalconditions}
\ee
Here we introduced the 2-site magnetization operators of the even and odd spin sub-lattices
\be
M_{a b,\iota}:= \sum_{\tau\in\mathbb Z_t+\frac{1}{2}\iota} \!\!\!\sigma_{a, \tau} \sigma_{b,\tau+\frac{1}{2}}\,, \qquad\qquad \iota\in\{0,1\}\,.
\label{eq:doubleM}
\ee
\end{lemma}
As a corollary of Lemma~\ref{lemma1} and Lemma~\ref{lemma2}, we can express the SFF in the limit $L\to\infty$
in terms of the dimension of the eigenspace of eigenvalue 1, which in turn (Lemma~\ref{lemma2})
equals the dimension (in ${\rm End}({\cal H}_{2t})$) of the commutant $\mathcal M'$ of the set 
\be
\mathcal M:= \{ M_{a,\iota} \}_{a,\iota} \cup \{ M_{ab,\iota} \}_{a,b,\iota}\,.
\label{eq:finalset}
\ee
Namely,
\be
\lim_{L\to\infty} K(t,L) = \dim \mathcal M'.
\label{eq:limitcommutant}
\ee
In fact, $\mathcal M'$ can be completely characterised:
\begin{theorem}
\label{theorem1}
The commutant $\mathcal M'$ is the span of the representation of the cyclic group $C_t$ of even-site translations on a periodic chain of $2t$ spins:
\be
\mathcal M' = {\rm span}\{ \Pi_{2t}^{2\tau};\,\, \tau = 0,1,\ldots t-1\}\,.
\label{eq:commutant}
\ee
\end{theorem}
Hence, we arrive at the following corollary, which summarises our first main result

\begin{corollary}
For local quantum circuits \eqref{eq:Floquet} with local gates of the form (\ref{eq:Floquetgates1},\ref{eq:Floquetgates2}) and  
\be
U = (u_1\otimes u_2) S e^{i J s_3 \otimes s_3} (u_3\otimes u_4),\qquad W = (u'_1\otimes u'_2) S e^{i J' s_3 \otimes s_3} (u'_3\otimes u'_4)
\ee
where $u_j, u'_j\in {\rm U}(d)$ and $J, J'\neq 0$, the SFF \eqref{eq:SFF} averaged according to the measure \eqref{eq:measure} fulfils 
\be
\lim_{L\to\infty} K(t,L) = t\,.
\label{eq:mainresult}
\ee
\end{corollary}
This is precisely the CUE result for \emph{all} times. It is remarkable that 2-point spectral correlations of dual-unitary circuits agree with RMT at all scales. 

Note that the restriction $J, J'\neq 0$ for the validity of the statement is not surprising. Indeed, for ${J=0}$ the gate $U$ does not encode interactions among the qudits, they are evolved in an entirely independent fashion (an analogous conclusion holds concerning $W$ for ${J'=0}$). This means that if one of $J$ and $J'$ is equal to zero not all the qudits are coupled by the dynamics and one cannot expect $\mathbb U_{L}$ in Eq.~\eqref{eq:Floquet} to behave like a random matrix on the whole Hilbert space.  

\subsection{Results on SFF at large $L$ for \emph{T-symmetric} dual-unitary circuits}

To obtain circuits with time-reversal symmetry one has to choose local gates $U, W \in {\rm U}(d^2)$ and on-site disorder $u_{x}, w_{x} \in {\rm U}(d)$ which are compatible with the conditions \eqref{eq:timereversal}. A generic choice is 
\be
U=U^T,\qquad\qquad W=W^T,
\label{eq:TIEc1}
\ee
and gates $u_{x}, w_{x}$ of the form 
\be
u_{x} = e^{i \boldsymbol{\theta}_{x}\cdot\boldsymbol{\sigma}},
\qquad
w_{x} = e^{i \boldsymbol{\theta}_{x}\cdot\boldsymbol{\sigma}^T} = u_x^T,
\qquad 
x\in\Lambda_L,\;\;\boldsymbol{\theta}_{x}\in\mathbb R^{d^2-1}.
\label{eq:urxT}
\ee
where now $\boldsymbol{\theta}_{x}$ is the same in both $u_{x}$ and $w_{x}$. The expectation can again be explicitly written in terms of a factorised measure as:
\be
\mathbb E_T[f] = \int  f(\boldsymbol\theta)  \prod_{x=0}^{L-1}\prod_{\iota'=0}^1
g_{\iota'}(\boldsymbol{\theta}_{x+\frac{\iota'}{2}}) {\rm d}^{d^2-1}\boldsymbol{\theta}_{x+\frac{\iota'}{2}}
\,,
\quad
\boldsymbol \theta \equiv (\boldsymbol{\theta}_{x})_{x\in\Lambda_L}\,.
\label{eq:measureT}
\ee 
where $g_{\iota'}\in L^1[\mathbb R^{d^2-1}]$ is a pair of arbitrary probability densities of i.i.d. random variables $\boldsymbol{\theta}_{x}$ on integer ($\iota'=0$) and half-odd-integer ($\iota'=1$) sublattices.

Considering local gates fulfilling the conditions (\ref{eq:TIEc1}, \ref{eq:urxT}) and averaging according to the measure $\mathbb E_T[\cdot] $, one can repeat the reasoning of Sec.~\ref{sec:duality} and conclude 
\bea
K_T(t,L) =  \mathbb E_T\left[\left({\rm tr}\,\mathbb U_{L}^t\right) \left({\rm tr}\,\mathbb U_{L}^t\right)^*\right]  = {\rm tr}\,\mathbb T_T^L,
 \label{eq:SFFdualityT}
\eea
with 
\be
\mathbb T_T : = (\tilde{\mathbb U} \otimes \tilde{\mathbb U}^*) {\mathbb O}_{T,1}^\dag (\tilde{\mathbb W} \otimes \tilde{\mathbb W}^*) {\mathbb O}^{\phantom{\dag}}_{T, 0}\,.
\ee
Here $\tilde{\mathbb U}$, and $\tilde{\mathbb W}$ are defined as in Eqs.~(\ref{eq:Utildee},\,\ref{eq:Utildeo}) while time-reversal symmetric averaging operator ${\mathbb O}_{T, \iota'}$ reads as     
\be
{\mathbb O}_{T, \iota'} := \int\!{\rm d}^{d^2-1}\boldsymbol{\theta}\,g_{\iota'}(\boldsymbol{\theta})
\exp\left(i\boldsymbol\theta\cdot (\boldsymbol{M} \otimes \1_{2t}-\1_{2t}\otimes \boldsymbol{M}^*)\right)\,.
\label{eq:OopT}
\ee
Here $\boldsymbol{M} = (M_{1},M_{2},\ldots, M_{d^2-1})$ with $M_{a}$ denoting the representation of the $a$-th Hermitian generator of ${\mathfrak su}(d)$ in the \emph{full} time lattice
\be
M_{a} :=\sum_{\tau \in \Lambda_t} \sigma_{a,\tau} = M_{a,0}+M_{a,1}.
\label{eq:fulllatticeM}
\ee
Assuming that, together with the conditions (\ref{eq:TIEc1}, \ref{eq:urxT}), the local gates also fulfil (\ref{eq:dualunitarityU}, \ref{eq:dualunitarityW}) (i.e. they are \emph{dual-unitary}), 
and noting that the averaging operator also satisifies the properties (\ref{eq:propi},\ref{eq:propii}),
one can immediately write an analogue of the Lemma~\ref{lemma1} for the transfer matrix $\mathbb T_T$ (its proof \ref{sec:prooflemma1} carries over). Namely, one can prove that $\mathbb T_T$ is again a linear non-expansive mapping over $\mathcal H_{2t}\otimes \mathcal H_{2t}$, and its eigenvectors corresponding to unimodular eigenvalues are determined by the conditions \eqref{eq:conditionsstateA} with $M_{a,\iota}$ replaced by $M_{a}$. Following the logic of Sec.~\ref{sec:leadingeigs}, one can map the problem of finding all linear independent eigenvectors of $\mathbb T_T$ corresponding to unimodular eigenvalues to a simpler algebraic problem. In this case the equivalent algebraic problem is to find all linearly independent operators $A$ fulfilling   
\be
\tilde{\mathbb U} \tilde{\mathbb W} A \tilde{\mathbb W}^\dag\,\tilde{\mathbb U}^\dag =e^{i \phi} A,\quad [M_{a},A] = 0, \quad [\tilde{\mathbb W}^\dag M_{a} \tilde{\mathbb W},A]=0\,.  \label{eq:CondSU2T}
\ee
for some $\phi\in\mathbb R$ and all $a\in\{1,2,\ldots,d^2-1\}$. 

Using the explicit from \eqref{eq:parametrisationgen} of the 2-site dual-unitary gates we can again simplify the conditions~\eqref{eq:CondSU2T} and write their invariant eigenoperator spaces in terms of simple algebraic commutants. In this case, to lift some technical complications, we specialise the treatment to the case of qubits ($d=2)$:
\begin{lemma}
\label{lemma3}
For $J, J'\neq 0$ and $d=2$ the conditions \eqref{eq:CondSU2T} cannot be met unless $\phi=0$. In this case, they are equivalent to 
\be
[A, M_{a}]=0,\; [A, M_{ab, \iota}+R_{2t} M_{ab, \iota}R_{2t}]=0\,, \quad a,b\in\{1,2,3\}\,, \; \iota\in\{0, 1\}\,,
\label{eq:finalconditionsT}
\ee
where the 2-site magnetization operators of the even and odd spin sub-lattices are defined in \eqref{eq:doubleM} and $R_{2t}$ is the reflection of the time lattice $\Lambda_t$ around the centre, i.e. 
\be
R_{2t} \ket{j_1}\otimes \ket{j_2}\otimes\cdots\ket{j_{2t}} = \ket{j_{2t}} \otimes \cdots\ket{j_2} \otimes \ket{j_{1}}.
\ee 
\end{lemma}
As before, this lemma allows us to express the SFF in the limit $L\to\infty$ in terms of the dimension (in ${\rm End}({\cal H}_{2t})$) of the commutant $\mathcal M'_T$ of the set 
\be
\mathcal M_T := \{ M_{a} \}_{a} \cup \{  M_{ab,\iota} +R_{2t} M_{ab, \iota}R_{2t} \}_{a,b,\iota}\,.
\ee
Namely,
\be
\lim_{L\to\infty} K_T(t,L) = \dim \mathcal M_T'.
\ee
The commutant $\mathcal M_T'$ can again be completely characterised by proving a statement analogous to Theorem~\ref{theorem1}, which we present here without proof. 
\begin{conjecture}
\label{conjecture1}
The commutant $\mathcal M'_T$ is the linear span of the representation of the dihedral group $D_t$, i.e. the symmetry group of a polygon of $t$ vertices, on a periodic chain of $2t$ spins:
\be
\mathcal M'_T = {\rm span}\{R_{2t}^n\Pi_{2t}^{2\tau};\,\, \tau = 0,1,\ldots t-1,\, n= 0,1\}\,.
\label{eq:commutantT}
\ee
\end{conjecture}
The number of independent elements of the dihedral group is established by:
\begin{lemma}
\label{lemma4}
The number of linearly independent elements in the representation of $D_t$ in $\mathcal H_t$ formed by $\{R_{2t}^n\Pi_{2t}^{2\tau}\}^{n=0,1}_{\tau=0,1,\ldots t-1}$ is $2t$. 
\end{lemma}
Hence, we arrive at the following corollary:  
\begin{corollary}
For local quantum circuits \eqref{eq:Floquet} with $d=2$, local gates of the form (\ref{eq:Floquetgates1},\ref{eq:Floquetgates2}), and  
\be
U = (u_1\otimes u_2) S e^{i J s_3 \otimes s_3} (u_1^{T}\otimes u_2^{T}),\quad W = (u'_1\otimes u'_2) S e^{i J' s_3 \otimes s_3} (u^{\prime\, T}_1\otimes u^{\prime\, T}_2),
\ee
where $u_j, u'_j\in {\rm U}(2)$ and $J, J'\neq 0$, the SFF \eqref{eq:SFF} averaged according to the measure \eqref{eq:measureT} fulfils 
\be
\lim_{L\to\infty} K_T(t,L) = 2t\,.
\ee
\end{corollary}
This is precisely the COE result for \emph{all} times. Once again we see that 2-point spectral correlations of dual-unitary circuits agree with RMT at all scales.

The proof of Conjecture~\ref{conjecture1} follows the same ideas as that of Theorem~\ref{theorem1} but is technically more involved and will be presented elsewhere~\cite{COE}. A similar proof, for the special case of the time-reversal symmetric self-dual kicked Ising model, has been presented in the supplemental material of Ref.~\cite{BKP18}.

\section{Proofs}
\label{sec:proofs}

\subsection{Proof of Lemma \ref{lemma1}}
\label{sec:prooflemma1}

 As a consequence of properties (\ref{eq:propi},\ref{eq:propii}) the spectra of $\mathbb O_{\iota\iota'}$ 
belong to the open unit disk with 1 attached to it, ${\mathcal D}_1 = \{z\in\mathbb C; |z|<1\} \cup \{1\}\supset {\rm spect}(\mathbb O_{\iota\iota'})$. 
Since $\mathbb O_{0\iota'}$ and $\mathbb O_{1\iota'}$ commute, ${\rm spect}(\mathbb O_{\iota'}=\mathbb O_{0\iota'}\mathbb O_{1\iota'}) \subset {\mathcal D}_1$, 
and all eigenvectors $\ket{R}$ of
$\mathbb O_{\iota'}$ with unique unimodular eigenvalue $1$ are characterised by
\be
(M_{a, \iota}\otimes \1_{2t}-\1_{2t}\otimes  M_{a, \iota}^*)\ket{R} = 0, \quad \iota\in\{0,1\},\;a\in\{1,2,\ldots,d^2-1\}\,.
\label{eq:unique}
\ee
Since $M_{a, \iota}\otimes \1_{2t}-\1_{2t}\otimes  M_{a, \iota}^*$ are Hermitian, exactly the same conditions uniquely characterise the eigenvalue $1$ eigenvectors of 
$\mathbb O^\dagger_{\iota'}$.

We now turn to SFF transfer matrix \eqref{eq:SFFTM2} and write
\be
\mathbb T^\dag \mathbb T=  \mathbb O_0^{{\dag}} (\tilde{\mathbb W} \otimes \tilde{\mathbb W}^*)^\dag \mathbb O_1^{\phantom{\dag}} \mathbb O_1^{{\dag}}   (\tilde{\mathbb W} \otimes \tilde{\mathbb W}^*) \mathbb O_0^{\phantom{\dag}}\,.
\label{eq:TT}
\ee
Let $\ket{A}$ be a normalised eigenvector of $\mathbb T$ associated to the eigenvalue $\lambda$. Considering the expectation value of \eqref{eq:TT} we 
have, since $\bra{B}\mathbb O^\dagger_{\iota'} \mathbb O_{\iota'}\ket{B} \le \langle B|B\rangle$, 
$\bra{B}\mathbb O_{\iota'} \mathbb O^\dagger_{\iota'}\ket{B} \le \langle B|B\rangle$, for any $\ket{B}$:
\bea
 |\lambda|^2 &=& \braket{A|\mathbb T^\dag \mathbb T|A} = \braket{A|\mathbb O_0^{{\dag}} (\tilde{\mathbb W} \otimes \tilde{\mathbb W}^*)^\dag \mathbb O_1 \mathbb O_1^{{\dag}} (\tilde{\mathbb W} \otimes \tilde{\mathbb W}^*) \mathbb O_0 |A}  \nonumber \\
&\le & \braket{A|\mathbb O_0^{{\dag}} (\tilde{\mathbb W} \otimes \tilde{\mathbb W}^*)^\dag (\tilde{\mathbb W} \otimes \tilde{\mathbb W}^*) \mathbb O_0 |A} = \braket{A|\mathbb O_0^{{\dag}} \mathbb O_0 |A} \leq 1\,, \label{eq:boundx}
\eea
which proves point (i). 

The eigenvalue $\lambda$ is unimodular only if both inequalities in (\ref{eq:boundx}) are saturated. 
The second one implies (\ref{eq:unique}) for $\ket{R}=\ket{A}$, i.e. the second line of \eqref{eq:conditionsstateA}, while the first one implies 
(\ref{eq:unique}) for $\ket{R}= (\tilde{\mathbb W} \otimes \tilde{\mathbb W}^*) \ket{A}$, i.e. the third line of \eqref{eq:conditionsstateA}.
Since $\mathbb O_0 \ket{A}=\ket{A}$, $\mathbb O^\dagger_1 (\tilde{\mathbb W} \otimes \tilde{\mathbb W}^*) \ket{A} = 
 (\tilde{\mathbb W} \otimes \tilde{\mathbb W}^*) \ket{A}$ we have the first line of \eqref{eq:conditionsstateA}. This proves point (ii).

Finally, we prove the last point by contradiction: assuming that the eigenvalue $\lambda$ corresponds to a non-trivial Jordan block, there must exist a normalised vector $\ket{B}$ such that
\be
\mathbb T \ket{B}=\lambda \ket{B} + \alpha \ket{A}\,,\qquad \alpha\neq0\,,
\ee 
where $\ket{A}$ is the eigenvector corresponding to the eigenvalue $\lambda$ (where we can choose $\braket{A|B}=0$). Reasoning as above we have
\be
\braket{B|\mathbb T^\dag \mathbb T|B} = 1+ |\alpha|^2 \ge 1\,,
\ee
which is a contradiction.

\subsection{Proof of Lemma \ref{lemma2}}

Plugging \eqref{eq:UGE} into the first condition \eqref{eq:CondSU2} and using the second condition to commute $\boldsymbol\alpha\cdot \boldsymbol M_{\iota}$ around $A$ we bring the conditions \eqref{eq:CondSU2} to the following form     
\begin{align}
\tilde{\mathbb V}\,\tilde{\mathbb V}^{\prime}  A \tilde{\mathbb V}^{\prime\dag} \tilde{\mathbb V}^\dag &\!\!=e^{i \phi} A, &  [M_{a,\iota},A] &\!=\!0, &  [\tilde{\mathbb V}^{\prime \dag} M_{a,\iota} \tilde{\mathbb V}^{\prime},A]&=0, \label{eq:intermediatecond}
\end{align}
where $a\in\{1,2,3\}$, $\iota\in\{0,1\}$.
Let us now consider more closely the operator in the last relation. Considering for example $\iota=0$
sublattice and a combination of generators which yields
the first spin matrix 
$s_1 = \boldsymbol \alpha \cdot \boldsymbol \sigma$,
\begin{align}
\tilde{\mathbb V}^{\prime \dag} (\boldsymbol\alpha\!\cdot\!\boldsymbol M_{0})\tilde{\mathbb V}^{\prime} &= \sum_{\tau \in\mathbb Z_t} \exp\left[- i J' s_{3, \tau-\frac{1}{2}} s_{3,\tau} \right]s_{1,\tau-\frac{1}{2}} \exp\left[i J' s_{3, \tau-\frac{1}{2}} s_{3,\tau} \right]\,,
\end{align}
where we used $S^\dag (\1\otimes s_a) S = s_a\otimes \1$. Resolving the identity in an eigenbasis of $s_{3}$ we find  
\begin{align}
\tilde{\mathbb V}^{\prime \dag} (\boldsymbol\alpha\!\cdot\!\boldsymbol M_{0})\tilde{\mathbb V}^{\prime} 
 &= \sum_{\tau\in\mathbb Z_t} s_{1,\tau-\frac{1}{2}} \cos(J's_{3,\tau}) + s_{2,\tau\!-\!\frac{1}{2}} \sin(J' s_{3, \tau}).
\label{eq:VMVe}
\end{align}
Next, we consider  
\begin{align}
&\tilde{\mathbb V}^{\prime \dag} (\boldsymbol\alpha\!\cdot\!\boldsymbol M_{0}) \tilde{\mathbb V}^{\prime} - e^{i\frac{\pi}{2} \boldsymbol\alpha\cdot\boldsymbol M_{0}} 
\tilde{\mathbb V}^{\prime \dag}  (\boldsymbol\alpha\!\cdot\!\boldsymbol M_{0}) \tilde{\mathbb V}^{\prime} e^{-i\frac{\pi}{2}\boldsymbol\alpha\cdot\boldsymbol M_{0}}= \notag\\
& = 2\!\sum_{\tau\in\mathbb Z_t}\! s_{2,\tau-\frac{1}{2}} \sin(J' s_{3, \tau}).\label{eq:VM1Vod}
\end{align}
Since $\sin(J' s_{3})$ is Hermitian and traceless it can be expanded in terms of the generators $\{\sigma_a\}$, i.e. 
\be
\sin(J' s_{3}) = \boldsymbol{c}(J')\cdot \boldsymbol\sigma\,,
\quad\textrm{where}\quad
\boldsymbol{c}(J') \neq \boldsymbol 0
\;\;\textrm{for}\;\; J'\neq 0\,.
\ee 
Furthermore, since the adjoint representation of 
${\rm SU}(d)$ is irreducible, we can for any non-vanishing vector
$\boldsymbol{\beta}\in\mathbb R^{d^2-1}$, and 
$b \in\{1,\ldots,d^2-1\}$
find a vector $\boldsymbol\gamma\in\mathbb R^{d^2-1}$, such that
\be
e^{i \boldsymbol\gamma\cdot\boldsymbol{M}_\iota}
(\boldsymbol{\beta}\cdot\boldsymbol{\sigma}_
{\tau+\frac{1}{2}\iota})
e^{-i \boldsymbol\gamma\cdot\boldsymbol{M}_\iota}=
\sigma_{b,\tau+\frac{1}{2}\iota},\quad
\tau\in\mathbb Z_t\,.
\ee
This means that, conjugating the operators on r.h.s. 
of \eqref{eq:VM1Vod} with appropriate 
$\boldsymbol\gamma\cdot\boldsymbol{M}_\iota$
on
integer ($\iota=0$) and half-odd integer ($\iota=1$) spin sub-lattices independently, we can produce any operator of the form $M_{a b,1}$  (cf.~\eqref{eq:doubleM}). Since $A$ commutes with $M_{a,\iota}$, we have for $J'\neq 0$:
\be
[A,M_{ab,1}]=0,\qquad  a,b\in\{1,2,\ldots,d^2-1\}.
\label{eq:AdoubleModd}
\ee 
To obtain an analogous statement for $M_{ab,0}$ we first note that combining the first and last relation of \eqref{eq:intermediatecond} yields  
\be
[\tilde{\mathbb V} M_{a,0} \tilde{\mathbb V}^\dag,A] =0\,, \qquad a\in\{1,2,\ldots, d^2-1\}\,.
\ee
Proceeding as before we find 
\be
\tilde{\mathbb V} (\boldsymbol\alpha\!\cdot\!\boldsymbol M_{1}) \tilde{\mathbb V}^{\dag} = \sum_{\tau\in\mathbb Z_t}   \cos(J s_{3,\tau}) s_{1,\tau+\frac{1}{2}}-\sin(J s_{3, \tau})s_{2,\tau+\frac{1}{2}},
\label{eq:VMVo}
\ee
and
\begin{align}
\!\!\!\!\!\! e^{i\frac{\pi}{2} \boldsymbol\alpha\cdot\boldsymbol M_{1}} 
\tilde{\mathbb V}
(\boldsymbol\alpha\!\cdot\!\boldsymbol M_{1}) \tilde{\mathbb V}^{\dag} e^{-i\frac{\pi}{2}\boldsymbol\alpha\cdot\boldsymbol M_{1}}- \tilde{\mathbb V} (\boldsymbol\alpha\!\cdot\!\boldsymbol M_{1}) \tilde{\mathbb V}^{\dag} & =  2\!\sum_{\tau\in\mathbb Z_t}\!\sin(J s_{3, \tau}) s_{2,\tau+\frac{1}{2}}.
\end{align}
Assuming $J\neq0$, we can repeat the reasoning after \eqref{eq:VM1Vod} and find:
\be
[A,M_{ab,0}]=0,\qquad  a,b\in\{1,2,\ldots, d^2-1\}.
\label{eq:AdoubleMeven}
\ee
Now we note that $\mathbb V$ and $\mathbb V'$ can be written in terms of double magnetisations~\eqref{eq:doubleM}. This can be seen by observing that 
\be
\sum_{a=1}^{d^2-1} \sigma_a\otimes\sigma_a
\ee 
is the quadratic Casimir operator of the representation of ${\rm SU}(d)$ over $\mathbb C^d \otimes \mathbb C^d$. Therefore, we must have 
\be
\sum_{a=1}^{d^2-1} \sigma_a\otimes\sigma_a = c_+ \frac{\1+S}{2} + c_- \frac{\1-S}{2} ,
\ee
for some $c_\pm\in\mathbb R$. Indeed, the symmetric and antisymmetric subspaces of $\mathbb C^d \otimes \mathbb C^d$ contain irreducible representations. Fixing the constants using the explicit form of $\{\sigma_a\}$ (see, e.g., Ref.~\cite{GenGellMann}) we find $c_\pm=\pm 2-2/d$. Using
\be
S = e^{- i \frac{\pi}{2}}e^{i \frac{\pi}{2} S},
\ee
and writing $s_3=\sum_a \gamma_a \sigma_a$
we finally find 
\begin{align}
\mathbb V &= e^{- i \frac{\pi}{2} t}\exp\Bigl[i \frac{\pi}{2} \sum_{\tau\in\mathbb Z_t} S_\tau\Bigr] \exp\Bigl[{i J\sum_{ab}\gamma_a\gamma_b M_{ab,0}}\Bigr] \notag\\
&= e^{- i \frac{\pi}{2d} t (1-d)}\exp\Bigl[{i \frac{\pi}{4} \sum_{a=1}^{d^2-1} M_{aa,0}}\Bigr] \exp\Bigl[{i J \sum_{ab} \gamma_a\gamma_b M_{ab,0}}\Bigr]\,, \label{eq:VMMMd}\\
\mathbb V' &= e^{- i \frac{\pi}{2} t} \Pi_{2t}\exp\Bigl[i \frac{\pi}{2} \sum_{\tau\in\mathbb Z_t} S_\tau\Bigr] \exp\Bigl[{i J' \sum_{a,b} \gamma_a\gamma_b M_{ab,0}}\Bigr] \Pi^\dag_{2t} \notag\\
&= e^{- i \frac{\pi}{2d} t (1-d)}\exp\Bigl[{i \frac{\pi}{4} \sum_{a=1}^{d^2-1} M_{aa,1}}\Bigr] \exp\Bigl[{i J'\sum_{ab}\gamma_a\gamma_b M_{ab,1}}\Bigr]\,, \label{eq:V'MMMd}
\end{align}
where we introduced 
\be
S_\tau :=  \1_{2\tau-1}\otimes S \otimes\1_{2t-2\tau-1}. 
\ee
Equations \eqref{eq:AdoubleModd}, \eqref{eq:AdoubleMeven}, \eqref{eq:VMMMd} and \eqref{eq:V'MMMd} imply that if $J,J'\neq0$
\begin{align}
\tilde{\mathbb V}\,\tilde{\mathbb V}^{\prime}  A \tilde{\mathbb V}^{\prime \dag} \tilde{\mathbb V}^\dag = A\,,
\end{align}
so that the first of conditions \eqref{eq:intermediatecond} can be fulfilled only for $\phi=0$ and in that case it follows from \eqref{eq:AdoubleModd}, \eqref{eq:AdoubleMeven}, and the second of \eqref{eq:intermediatecond}. This proves the Lemma. 

\subsection{Proof of Theorem \ref{theorem1}}

To prove the Theorem we use of the following Lemma. 

\begin{lemma}
\label{lemma5}
Let $\mathcal K \subseteq {\rm End}(\mathcal H_{2t})$ be a multiplicative algebra of operators over $\mathcal H_{2t}$, generated by 
elements of $\mathcal M = \{ M_{a,\iota} \}_{a,\iota} \cup \{ M_{ab,\iota} \}_{a,b,\iota}$.
The representations of $\mathcal K$ over the eigenspaces $\{\mathcal W_k \subset \mathcal H_{2t} \}_{k=0}^{t-1}$ of the 2-site shift operator $(\Pi_{2t})^2\mathcal W_k = e^{-2\pi i k/t}\mathcal W_k$ are all irreducible and inequivalent.
\end{lemma}

Our statement (Theorem~\ref{theorem1}) follows from a simple combination of Lemma~\ref{lemma5} (which is proven later below) and the Schur's Lemma. 
 If some $A$ fulfils the conditions \eqref{eq:finalconditions}, it commutes with all elements of the algebra $\mathcal K$ generated by 
 $\{M_{a, \iota}\}$ and $\{M_{ab, \iota}\}$. Since the representations of the algebra $\mathcal K$ in the eigenspaces $\mathcal W_k$ are irreducible and inequivalent, Schur's Lemma implies  
\be
A = \sum_{k=0}^{t-1} c_k Q_k, \label{eq:sol}
\ee
where $c_k\in\mathbb C$ are arbitrary coefficients and 
\be
Q_k := \frac{1}{t}\sum_{\tau=0}^{t-1} e^{2\pi i \tau k/t} \Pi_{2t}^{2\tau}\,,\qquad k\in\{0,\ldots,t-1\},
\label{eq:Zk}
\ee
are orthogonal projectors on $\{\mathcal W_k\}_{k=0}^{t-1}$, i.e. $Q_k Q_{k'} = \delta_{k,k'} Q_k$. 
This proves that $A$ is a linear combination of $t$ cyclic translations $\Pi_{2t}^{2\tau}$, i.e. $\mathcal K' = \mathcal M' = 
{\rm span}\{\Pi^{2\tau}_{2t};\,\tau=0,1\ldots,t-1\}$, concluding the proof of Theorem \ref{theorem1}.

\subsection{Proof of Lemma \ref{lemma5}}

We begin by introducing the following shorthand notation
\begin{align}
S_{\pm}& := \sum_{\tau\in \Lambda_t}
s_{\pm,\tau} \left(s_{3,\tau}+\frac{d-1}{2}\right)\,,\label{eq:Spm} \\
R_{\pm, n}& := \sum_{\tau\in \Lambda_t} 
\left(s_{3,\tau}+\frac{d-1}{2}\right) s^n_{\pm,\tau+\frac{1}{2}}\,,\label{eq:Rpm} \\
T_{\pm, n}& := \sum_{\tau\in \Lambda_t} 
\left(s_{3,\tau}-\frac{d-1}{2}\right) s^n_{\pm,\tau-\frac{1}{2}}\,,\label{eq:Tpm} \\
M_{\pm\mp}& := \sum_{\tau\in \Lambda_t} s_{\pm,\tau}s_{\mp,\tau+\frac{1}{2}},\label{eq:Mpmmp} \\
Z_\iota & := \sum_{t\in\mathbb Z_t}
s_{3,\tau+\frac{1}{2}\iota},\label{eq:Zdef}
\end{align}
where $n\in\{1,\ldots,d-1\}$ and 
\be
s_{\pm,\tau} = \frac{s_{1,\tau} \pm i s_{2,\tau}}{\sqrt 2},\qquad \tau\in\Lambda_t,
\ee
are local spin raising/lowering operators. Using the commutation relations among $\{\sigma_a\}$ it is straightforward to show that all operators \eqref{eq:Spm}--\eqref{eq:Zdef} can be expressed as linear combinations of $M_{a,\iota}$, $M_{ab,\iota}$ (cf.~\eqref{eq:sublatticeM} and \eqref{eq:doubleM}). 

In addition, we also introduce the set of vectors (states) in $\mathcal H_{t}$
\be
\!\!\!\!\!\mathcal S_k : =\!\! 
\begin{cases}
\begin{aligned}
& \ket{0}^{\otimes 2t} \cup \{ \ket{n}_0;\, n\in \mathcal I\}\,\cup \\
&\{\ket{n, {\boldsymbol\nu}, {m}}_{\ell, 0};\, n, m \in \mathcal I,  \boldsymbol\nu\in \mathcal J^{\ell-2},
2\le\ell\le 2t\}
\end{aligned}
 & k=0\\
 \\
\begin{aligned}
&\{ \ket{n}_k;\, n\in \mathcal I\}\,\cup \\
&\{\ket{n, {\boldsymbol\nu}, {m}}_{\ell, k};\, n, m \in \mathcal I, \boldsymbol\nu\in \mathcal J^{\ell-2},
2\le\ell\le 2t\}
\end{aligned}
 & k\!\in\!\{1,\ldots,\!2t\!-\!1\!\}
\end{cases}
\label{eq:states}
\ee
where we defined the sets
\begin{align}
 \mathcal I &:= \{1,\ldots,d-1\},\label{eq:setI}\\
 \mathcal J &:= \{0, 1,\ldots,d-1\},
\end{align}
and the vectors 
\begin{align}
&\ket{n}_{k}:=\frac{1}{\sqrt{2t}}\sum_{j=0}^{2t-1} e^{i\frac{\pi k}{t} j}\Pi^j_{2t} s^{n}_{+,0} \ket{0}^{\otimes 2t},\label{eq:statesm1}\\
&{\ket{n_1, \underbrace{0\cdots0}_{\ell_1-1}, n_{2}, \cdots, n_a, \underbrace{0\cdots0}_{\ell_a-1}, n_{a+1}}\!{}_{\ell, k}}\notag\\
&\qquad\qquad := \frac{1}{\sqrt{2t}}\sum_{j=0}^{2t-1} e^{i\frac{\pi k}{t} j}\Pi^j_{2t} s^{n_1}_{+,0}s_{+,\frac{\ell_1}{2}}^{n_{2}}\!\!\cdots s_{+,\frac{\ell-\ell_a}{2}}^{n_{a}}s_{+,\frac{\ell-1}{2}}^{n_{a+1}} \ket{0}^{\otimes 2t}\!\!,\label{eq:statesm>1}
\end{align}
with $\{\ell_j\}_{j=1}^a\subset\{1,\ldots, 2t-1\}$ fulfilling
\be
\mathbb N \ni \ell :=1+\sum_{j=1}^a\ell_{j}\leq 2t\,.
\ee
We note that the `empty' state $\ket{0}\in \mathcal R$ (cf. \eqref{eq:realbasis}) 
satisfies 
\be
s_3\ket{0}= -\frac{d-1}{2}\ket{0},
\ee
(cf. \eqref{eq:diag}). The integer $\ell$ shall be referred to as the \emph{length} of the states~\eqref{eq:statesm>1} (one can verify that $\ell\geq 2$), while the states \eqref{eq:statesm1} have conventionally a unit length. For each value of $k$ the set \eqref{eq:states} contains $(d-1)^2d^{\ell-2}$ states for every length $\ell\geq2$ and $d-1$ states for $\ell=1$.

Note that for each $k$ the states in \eqref{eq:states} have 
\emph{momentum} $\pi k/t$, i.e.
\be
\Pi_{2t} \ket{n, {\boldsymbol \nu}, m}_{\ell, k} = e^{-i \frac{\pi k}{t}} \ket{n, {\boldsymbol \nu}, m}_{\ell, k}\,. 
\ee 
The set $\mathcal S_k$ is complete in $\mathcal V_k$ --- the eigenspace of single-site shift $\Pi_{2t}$ associated with momentum $\pi k/t$ --- but are  not all linearly independent: 
While for $\ell < t$ the states are clearly orthonormal, some of the states with $\ell \geq t$ can be represented by a string with a shorter length or they have multiple representations with the same length. 
One can then construct a basis $\mathcal B_k$ of $\mathcal V_k$ by extracting from $\mathcal S_k$ the maximal subset of linearly independent vectors. 

Here we want to prove the following (Lemma \ref{lemma5}):
The representation of the algebra $\mathcal K$ generated by $\{M_{a,\iota}\}$ and $\{M_{ab,\iota}\}$ is irreducible in 
\be
\mathcal W_k\equiv{\rm span}(\mathcal B_k \cup \mathcal B_{k+t}),\qquad\qquad k\in\{0,1,\ldots,t-1\}\,, 
\ee
specifically in the eigenspace of $(\Pi_{2t})^2$ corresponding to the eigenvalue $e^{-{2\pi i k}/{t}}$. Moreover, the irreducible representations in different $\mathcal W_k$ are inequivalent.

Noting that $\mathcal W_k$ are closed under the action of $\mathcal K$ (all generators commute with $(\Pi_{2t})^2$) we have that the following three requirements 
imply the statement of the Lemma:
\begin{itemize}
\item[(1)]  All vectors in $\mathcal B_k$ are mapped into one another by elements of the algebra $\mathcal K$. 
\item[(2)] There is an element of $\mathcal K$ mapping $\ket{1}_k$ and $\ket{1}_{k+t}$ one into another. 
\item[(3)]  There is no unitary matrix $C$ such that
\begin{align}
(Z_1)_k &= C (Z_1)_p C^{\dag},\label{eq:condC1}\\
(Z_0)_k &= C (Z_0)_p C^{\dag}, \label{eq:condC2}\\
(M_{+-}^2)_k &= C (M_{+-}^2)_p C^{\dag},\label{eq:condC3}
\end{align}
where $(\cdot)_p$ denotes the projection to $\mathcal W_p$, if $p\neq k$.
\end{itemize}

\noindent\emph{Proof of (1)}. We prove the statement by showing the validity of a sufficient condition: all states \eqref{eq:states} are mapped into one-another by elements of the algebra $\mathcal K$. This condition is sufficient because the elements of $\mathcal B_k$ are a subset of the states \eqref{eq:states}.  

We begin by proving that one can map $\ket{1}_k$ into every state \eqref{eq:states}. First, we note that using 
\be
S_{\pm} \ket{n}_{k} = n \ket{n\pm1}_{k}
\label{eq:Spmrel}
\ee
we can map $\ket{1}_{k}$ to all states $\ket{n}_{k}$ with $n\in\{2,\ldots,d-1\}$ and, for $k=0$, also to $\ket{0}^{\otimes 2 t}$. Next, we observe that using 
\be
R_{+,1} \ket{n}_{k} = n \ket{n, 1}_{2,k}\label{eq:Rprel}
\ee
and then repeatedly applying 
\be
S_{+} \ket{n,m}_{k} = n \ket{n+1, m}_{2,k}+ m \ket{n, m+1}_{2,k}
\ee
we can map $\ket{1}_{k}$ into every state $\ket{n, m}_{2,k}$ with $n,m\in\{1,\ldots,d-1\}$. 

We proceed using an inductive argument. Assuming that we can access every state $\ket{m, {\boldsymbol \nu}, m}_{\ell', k}$ of length $\ell'<\ell$ we shall prove that we can access every state of length
$\ell$, for $\ell \geq 3$. This follows straightforwardly from the relations  
\begin{align}
R_{+,1} \ket{n, {\boldsymbol\nu}, m}_{\ell-1, k} &\simeq m \ket{n, {\boldsymbol\nu}, m, 1}_{\ell, k}\,,\label{eq:indstep1}\\
M_{-+} \ket{{n, {\boldsymbol\nu}, 1}}_{\ell-1,k} &\simeq \ket{n,{\boldsymbol\nu},0,1}_{\ell, k} \,,\label{eq:indstep2}
\end{align}
where $\simeq$ denotes equality up to states of length $<\ell$ which can be accessed by assumption, and the repeated application of 
\begin{align}
S_{+} \ket{n, {\boldsymbol\nu}, m, m_2}_{\ell, k} &\simeq m_2 \ket{n, {\boldsymbol\nu}, m, m_2+1}_{\ell, k}\,,\label{eq:fe}\\
S_{+} \ket{n, {\boldsymbol\nu}, 0, m_2}_{\ell, k} &\simeq  m_2 \ket{n, {\boldsymbol\nu}, 0, m_2+1}_{\ell, k}\,. \label{eq:se}
\end{align}
In Eq.~\eqref{eq:fe}, $\simeq$ denotes equality up to states of the form $\ket{n', {\boldsymbol\nu}', m', m_2}_{\ell, k}$ that are accessed at the previous step ($n$, $\boldsymbol \nu$, and $m$ are arbitrary). Analogously in Eq.~\eqref{eq:se}, $\simeq$ denotes equality up to states of the form $\ket{n', {\boldsymbol\nu}', 0, m_2}_{\ell, k}$. 

This means that for every state $\ket{n, {\boldsymbol\nu}, m}_{\ell, k}$ there exist an operator $B_{n, \boldsymbol \nu, m}\in\mathcal K$ such that 
\be
B_{n, \boldsymbol \nu, m}\ket{1}_{k}= \ket{n, {\boldsymbol\nu}, m}_{\ell, k}\,.
\ee
Then we can construct an operator mapping the arbitrary vector $\ket{n', {\boldsymbol \nu}', m'}_{\ell', k}$ into the arbitrary vector $\ket{n, {\boldsymbol \nu}, m}_{\ell, k}$
\be
A_{n, \boldsymbol \nu, m; n', \boldsymbol \nu', m'}=B^{\phantom{\dag}}_{n, \boldsymbol \nu, m} \ket{1}_{\!k\,\,k}\!\!\bra{1} B^\dag_{n', \boldsymbol \nu', m'}\,.
\ee
To prove that this operator is in $\mathcal K$ we fist note that, since the generators are Hermitian, we have that if ${B^{\phantom{\dag}}_{n, \boldsymbol \nu, m}\in\mathcal K}$ also ${B^\dag_{n, \boldsymbol \nu, m}\in\mathcal K}$. We then just need to prove that $\ket{1}_{k}\prescript{}{k}{\bra{1}}\in\mathcal K$. This is explicitly done by observing   
\begin{align}
&\ket{1}_{k} \prescript{}{k}{\bra{1}}= \frac{(S_{-})^{d-2} (S_{+})^{d-2} T_{-,d-1}^{2t-1} R_{+,d-1}^{2t-1}}{\prescript{}{k}{\braket{1|(S_{-})^{d-2} (S_{+})^{d-2} T_{-,d-1}^{2t-1} R_{+,d-1}^{2t-1}|1}_k}}\,, &  &k=0 \lor d\neq 2\,,\\
&\ket{1}_{k} \prescript{}{k}{\bra{1}}= \frac{T_{-,1}^{2t-2} R_{+,1}^{2t-2}}{\prescript{}{k}{\braket{1|T_{-,1}^{2t-2} R_{+,1}^{2t-2}|1}_k}}\,, & &k\neq 0 \wedge d= 2\,.
\end{align}

\noindent\emph{Proof of (2)}. This point is immediate. Indeed, one can directly verify that 
\be
(Z_{0}-Z_{1}) \ket{1}_{k} = \ket{1}_{k+t}\,.
\ee

\noindent\emph{Proof of (3)}. The statement is trivial whenever $k$ or $p$ are 0. Indeed, $\ket{0}^{\otimes 2t}$ is the only eigenstate of $Z_0+Z_1$ corresponding to eigenvalue $2t$ and does not appear for $p\neq0$. This means that the sum of \eqref{eq:condC1} and \eqref{eq:condC2} can never be satisfied. 

To prove (3) for $k,p\neq0$ we note that 
\be
\ket{1}^{(0)}_{k} : = \frac{\ket{1}_{k}+\ket{1}_{k+t}}{\sqrt 2}\,,
\ee
is the only eigenstate of $Z_{1}$ and $Z_{0}$ with eigenvalues $t$ and $t-2$ respectively. This means that \eqref{eq:condC1} and \eqref{eq:condC2} can be fulfilled only if $\ket{1}^{(0)}_{k}$ is an eigenstate of $C$. In turn, this implies that 
\be
 \tensor*[^{(0)}_{k}]{\braket{1 |(M_{+-})^2| 1}}{_k^{(0)}} = e^{-2i \pi k/t}\,,
\ee
 is invariant under the mapping implemented by $C$. Since $e^{-2i \pi k/t}\neq e^{-2i \pi p/t}$ for $k\neq p \in \{1,\ldots, t-1\}$ we conclude that there can be no transformation $C$ fulfilling \eqref{eq:condC2}.

\subsection{Proof of Lemma \ref{lemma3}}

We consider $d=2$, where $s_a = \frac{1}{2}\sigma_a$, $a\in\{1,2,3\}$, and begin by writing the analogue of \eqref{eq:intermediatecond}. To this aim we plug the form \eqref{eq:parametrisationgen} in the definitions (\ref{eq:Utildee}, \ref{eq:Utildeo}) and use the constraints (\ref{eq:TIEc1}, \ref{eq:urxT}) to find 
\begin{align}
&\tilde{\mathbb U} = e^{i \theta} e^{i \boldsymbol{\alpha} \cdot
\boldsymbol M}\,\tilde{\mathbb V}\,
e^{i \boldsymbol\beta\cdot\boldsymbol M}, & &\tilde{\mathbb W} = e^{i \theta'} e^{i \boldsymbol\gamma\cdot\boldsymbol M}\,  \tilde{\mathbb V}'\,
e^{i \boldsymbol\delta\cdot \boldsymbol M}\!, \label{eq:UTIE}
\end{align}
where $\boldsymbol{\alpha},\boldsymbol{\beta},\boldsymbol{\gamma},\boldsymbol{\delta}\in\mathbb R^3$, $\theta,\theta'\in\mathbb R$,
$\boldsymbol M = (M_1,M_2,M_3)$, while
$\tilde{\mathbb V}$ and $\tilde{\mathbb V}'$ are defined in \eqref{eq:genmag}.
Substituting now \eqref{eq:UTIE} into the first condition \eqref{eq:CondSU2T} and using the second condition to commute $\boldsymbol\alpha\cdot\boldsymbol M$ around $A$ we find the desired analogue of \eqref{eq:intermediatecond}
\begin{align}
\tilde{\mathbb V}\,\tilde{\mathbb V}^{\prime}  A \tilde{\mathbb V}^{\prime \dag} \tilde{\mathbb V}^\dag &\!\!=e^{i \phi} A, &  [M_a,A] &\!=\!0, &  [\tilde{\mathbb V}^{\prime \dag} M_a \tilde{\mathbb V}^{\prime},A]&=0,   & a&\in\{1,2,3\}\,. \label{eq:intermediatecondTIE}
\end{align}
Next we consider 
\begin{align}
\tilde{\mathbb V}^{\prime \dag} M_1 \tilde{\mathbb V}' &=\tilde{\mathbb V}^{\prime \dag} M_{1,1} \tilde{\mathbb V}' +\tilde{\mathbb V}^{\prime \dag} M_{1,0} \tilde{\mathbb V}' \notag\\
&= \sin(2J') (M_{32,1}+M_{23,1}) + \cos(2J') M_{1},\\
\tilde{\mathbb V} M_1 \tilde{\mathbb V}^\dag &= \tilde{\mathbb V} M_{1,1} \tilde{\mathbb V}^\dag+\tilde{\mathbb V} M_{1,0} \tilde{\mathbb V}^\dag  \notag\\
&= - \sin(2J) (M_{32,0}+M_{23,0}) + \cos(2J) M_{1}\,.
\end{align}
where we used \eqref{eq:VMVe} and \eqref{eq:VMVo} specialised to the case $d=2$. 

From these relations we see that, for $J, J'\neq 0$, $A$ commutes with $\{M_{32, \iota}+M_{23, \iota}\}_{\iota=0,1}$. Using the second of \eqref{eq:intermediatecondTIE} to make generic ${\rm SU}(2)$ rotations we find that actually $A$ commutes with the following 10 operators 
\begin{align}
\{M_{12, \iota}+M_{21, \iota},&M_{13, \iota}+M_{31, \iota},\notag\\
&M_{23, \iota}+M_{32, \iota}, M_{11, \iota}-M_{22, \iota},M_{11, \iota}-M_{33, \iota}\}_{\iota=0,1}.
\label{eq:MsymTIE}
\end{align}
In fact, $A$ commutes also with $\{M_{aa,\iota}\}$. To show that we consider the following objects 
\begin{align}  
&P_{\bar \iota} (P_{\iota}P_{\bar \iota})^{\frac{t}{2}-1} (M_{11,\iota}-M_{33,\iota}) ({P_{ \iota}P_{\bar \iota}})^{1-\frac{t}{2}} P^{-1}_{\bar \iota}\notag\\ 
&= - S_3 M_{11,\iota} -M_{33,\iota}, \qquad\qquad\qquad\quad t\,\,\text{even}\label{eq:stringteven}\\
\notag\\
&(P_{\iota}P_{\bar \iota})^{\frac{t-1}{2}} (M_{11,\iota}-M_{33,\iota}) (P_{\iota}P_{\bar \iota})^{-\frac{t-1}{2}} \notag\\ 
&= - S_3 M_{11,\bar \iota} + M_{33,\bar \iota}, \qquad\qquad\qquad \quad t\,\,\text{odd}\label{eq:stringtodd}
\end{align}
for $\iota\in\{0,1\}$, $\bar{\iota}:=1-\iota$, and we defined 
\be
S_a = \sigma^{\otimes 2t}_a= i^{2t} e^{i \frac{\pi}{2} M_a}, \qquad\qquad P_{\iota} = e^{i \frac{\pi}{4} (M_{11,\iota}-M_{22,\iota})}\,.
\ee
The left hand sides of \eqref{eq:stringteven} and \eqref{eq:stringtodd} commute with $A$ by construction, therefore we find   
\begin{align}  
&[A, S_3 M_{11, \iota} +M_{33, \iota}]=0, & & t\,\,\text{even}\label{eq:stringteven1}\\
& [A, S_3 M_{11, \iota} - M_{33, \iota}]=0, & & t\,\,\text{odd}\label{eq:stringtodd1}.
\end{align}
Using that $A$ commutes with $\{S_a\}_{a=1,2,3}$ and with the operators \eqref{eq:MsymTIE} we then have 
\begin{align}  
&[A, (S_a+\1_t) M_{bb, \iota}]=(S_a+\1_t) [A,  M_{bb, \iota}]=0, & & t\,\,\text{even}\label{eq:stringteven2}\\
& [A, (S_a-\1_t) M_{bb, \iota} ]=(S_a-\1_t) [A,  M_{bb, \iota}]=0, & & t\,\,\text{odd}\,.\label{eq:stringtodd2}
\end{align}
At this point we observe
\be
{\rm spect}(S_1+S_2+S_3)= \{(-1)^{t} 3,(-1)^{t+1}\}\,,
\ee
where the multiplicities (both algebraic and geometrical) of the two eigenvalues are $2^{2t-2}$ and $3 \cdot 2^{2t-2}$ respectively. 

Thus, by summing (\ref{eq:stringteven2}, \ref{eq:stringtodd2}) over $a$ we obtain that the commutators are multiplied by invertible operators. Therefore, we finally arrive at   
\be
[A, M_{bb, \iota}]=0\,,\qquad b\in\{1,2,3\}\,.
\label{eq:AdoubleMevenTsym}
\ee
Putting it all together we find  
\be
[A, M_{ab, \iota} +R_{2t} M_{ab, \iota} R_{2t} ]=0\,.
\ee
At this point, considering $\tilde{\mathbb V}\,\tilde{\mathbb V}'$ and using (\eqref{eq:VMMMd}, \eqref{eq:V'MMMd}) we have 
\begin{align}
\tilde{\mathbb V}\,\tilde{\mathbb V}^{\prime}  A \tilde{\mathbb V}^{\prime \dag} \tilde{\mathbb V}^\dag = A\,,
\end{align}
so that the first of conditions \eqref{eq:intermediatecondTIE} can be fulfilled only for $\phi=0$, and in that case it follows from \eqref{eq:AdoubleMevenTsym}, and the second of \eqref{eq:intermediatecondTIE}. This proves the Lemma.  
 
\subsection{Proof of Lemma \ref{lemma4}}

To prove the statement we note that the set of operators $\{R_{2t}^n\Pi_{2t}^{2\tau}\}^{n=0,1}_{\tau=0,1,\ldots t-1}$ can be written as 
\be
\Pi_{2t}^{2\tau} = \sum_{k=0}^{t-1} e^{-2\pi i \tau k/t} Q_k,  \qquad\qquad R_{2t} \Pi_{2t}^{2\tau}= \sum_{k=0}^{t-1} e^{-2\pi i \tau k/t} Q'_k,
\ee
where $Q_k$ are the orthogonal projectors defined in Eq.~\eqref{eq:Zk} and we introduced 
\be
Q'_k := \frac{1}{t}\sum_{\tau=0}^{t-1} e^{2\pi i \tau k/t} R_{2t} \Pi_{2t}^{2\tau}\,, \qquad k\in\{0,\ldots,t-1\}.
\label{eq:Z'k}
\ee
Since the mapping between $\{R_{2t}^n\Pi_{2t}^{2\tau}\}^{n=0,1}_{\tau=0,1,\ldots t-1}$ and $\{Q_k, Q'_k\}_{k=0,1,\ldots t-1}$ is invertible it is sufficient to prove that the latter operators are linearly independent. To this aim we note that $\{Q_k, Q'_k\}_{k=1,\ldots t-1}$ are obviously linearly independent. This can be seen by writing them in a basis of eigenstates of $\Pi_{2t}$ and noting distinct operators are non zero on distinct, non-overlapping, blocks. Moreover, noting that all $\{Q_k, Q'_k\}_{k=1,\ldots t-1}$ are zero when reduced to the block of zero two-momentum (i.e. the one composed by eigenstates of $\Pi_{2t}$ with eigenvalues $1$ and $-1$), we have that the only two operators which can be linearly dependent are $Q_0$ and $Q'_0$. To prove that such operators are independent we note that 
\be
Q_0 \ket{n}_0 = Q'_0 \ket{n}_0 = \ket{n}_0\,,\quad Q_0 \ket{n}_t = -Q'_0 \ket{n}_t = \ket{n}_t\,,\quad n\in\mathcal I\,,
\ee
where $\ket{n}_0, \ket{n}_t$ are defined in Eq.~\eqref{eq:states} and the set $\mathcal I$ in \eqref{eq:setI}. This implies that
\be
\alpha Q_0+\beta Q'_0 =0
\ee
only if $\alpha=\beta=0$ and concludes the proof.

\section{Discussion of the results and their possible extensions}
\label{sec:discussion}

In this section we discuss some generalisations and extensions of our results. While the extension to inhomogeneous interactions in Sec.~\ref{Sec:inhom} is rigorous, the other two subsections discussing fluctuations of SFF (\ref{sec:fluctuations}) and singular disorder distributions (\ref{sec:nonisotropic}) are currently of speculative nature.

\subsection{Spatially inhomogeneous interactions}
\label{Sec:inhom}

The space-time duality approach adopted in this paper treats separately each point in space and it is therefore convenient to study general inhomogeneous interactions. However, our results as elaborated in Sec.~\ref{sec:main} are not directly applicable to this case. Here we explain how to extend them focussing for simplicity on the case of no time reversal symmetry.  

We begin by observing that for position-dependent local gates, Eq.~\eqref{eq:SFFduality} is substituted with  
\be
K(t,L) = {\rm tr}\left(
\mathbb T_1 \mathbb T_2 \cdots 
\mathbb T_L\right)\,,
\label{eq:inhomoK}
\ee
where $\mathbb T_x$ is defined as in Eq.~\eqref{eq:SFFTM2} with $\mathbb O$ given in Eq.~\eqref{eq:defOa} while $\tilde{\mathbb U}$ and $\tilde{\mathbb W}$ are replaced by 
\begin{align}
\tilde{\mathbb W}_x &:= \prod_{\tau \in \mathbb Z_{t}} \eta_{\tau,t}(\tilde W_{x+\frac{1}{2}}),\\
\tilde{\mathbb U}_x &:= \!\prod_{\tau \in \mathbb Z_{t}+\frac{1}{2}}\!\eta_{\tau,t}(\tilde U_x).
\end{align}
Then we choose dual-unitary gates $U_x,W_x$ of the form \eqref{eq:parametrisationgen}, assuming 
$J_x, J_x' \neq 0$ for all $x$. We use Lemma~\ref{lemma1}, Lemma~\ref{lemma2}, and Theorem~\ref{theorem1} to find 
\be
K(t,L) = t + {\rm tr}\left(
\mathbb R_1 \mathbb R_2 \cdots 
\mathbb R_L\right)\,,
\label{eq:inhomoK2}
\ee
where $\mathbb R_{x}= (\1-\mathbb P) \mathbb T_x (\1-\mathbb P)$ and $\mathbb P$ is the projector onto the eigenspace of $\mathbb T_x$ associated to the eigenvalue 1. In writing \eqref{eq:inhomoK2} we used that, due to Lemma~\ref{lemma1}, Lemma~\ref{lemma2}, and Theorem~\ref{theorem1}, such eigenspace is the same for all $x$.

From \eqref{eq:inhomoK2} we see that to recover the result \eqref{eq:mainresult} we need to show that 
\be
\lim_{L\to\infty}
{\rm tr}\left(
\mathbb R_1 \mathbb R_2 \cdots 
\mathbb R_L\right) = 0\,.
\label{eq:dream}
\ee
For example this would hold if  $\| \mathbb R_{x}\| <1$, where $\|\cdot\|$ denotes the operator norm. The results of Sec.~\ref{sec:main}, however, are not sufficient to infer \eqref{eq:dream}.  

To overcome this problem, we focus on the case of $L$ even and make the following different replacement 
\be
\mathbb T_{x-1} \mathbb T_x =  \mathbb P + \tilde{\mathbb R}_{x-1, x} \quad \Rightarrow \quad K(t,L) = t + {\rm tr}(\tilde{\mathbb R}_{1,2} \tilde{\mathbb R}_{3,4} \cdots \tilde{\mathbb R}_{L-1,L})\,.
\label{eq:replacement}
\ee
where 
\be
\tilde{\mathbb R}_{x-1, x}= (\1-\mathbb P) \mathbb T_{x-1} \mathbb T_x (\1-\mathbb P).
\ee For the new operator $\tilde{\mathbb R}_{x-1, x}$ we are able to prove the following theorem (the proof is provided at the end of the subsection):
\begin{theorem}
\label{theorem2}
For $J_x,J'_x \neq0$, 
\be
\|\tilde{\mathbb R}_{x-1, x}\| < 1.
\ee
\end{theorem} 
This means that if there is a {\em finite density} of points $x$ with nonzero couplings ($J_x,J'_x\neq 0$), i.e. finite density of non-SWAP gates, then
\be
\lim_{L\to\infty}
{\rm tr}\left(
\tilde{\mathbb R}_{1,2}  \cdots 
\tilde{\mathbb R}_{L-1,L}\right) = 0\,,
\label{eq:dream2}
\ee
and 
\be
\lim_{L\to\infty} K(t,L) = {\rm tr}\,\mathbb P = t\,.
\ee
A completely analogous treatment holds for $L$ odd, e.g., considering  
\be
K(t,L) = t + {\rm tr}(\tilde{\mathbb R}_{12} \cdots \tilde{\mathbb R}_{L-2,L-1}{\mathbb R}_{L})\,.
\ee

\subsubsection{Proof of Theorem~\ref{theorem2}}
To prove the statement it is sufficient to show that if 
\be
\braket{A| (\mathbb T_{x-1} \mathbb T_x)^\dag \mathbb T_{x-1} \mathbb T_x|A}=1
\ee
for some state $\ket{A}$, then 
\be
\mathbb P \ket{A}= \ket{A}\,.
\ee
This follows immediately from Theorem~\ref{theorem1} and the following two lemmas
\setcounter{lemma}{5}
\begin{lemma}
\label{lemma6}
For dual-unitary circuits, if a state $\ket{A}$ fulfils 
\be
\braket{A|(\mathbb T_0 \mathbb T)^\dag\mathbb T_0 \mathbb T|A}=1
\label{eq:EV}
\ee
where $\mathbb T_0$ and $\mathbb T$ are transfer matrices of the form \eqref{eq:SFFTM2} (with unitary matrices $\tilde{\mathbb U}_0$ and $\tilde{\mathbb W}_0$ and $\tilde{\mathbb U}$ and $\tilde{\mathbb W}$ respectively), then 
\begin{align}
&
( M_{a, \iota}\otimes \1_{2t}-\1_{2t}\otimes  M_{a, \iota}^*) (\tilde{\mathbb W}_0 \otimes \tilde{\mathbb W}^*_0)  (\tilde{\mathbb U} \otimes \tilde{\mathbb U}^*) (\tilde{\mathbb W} \otimes \tilde{\mathbb W}^*)\ket{A}=0\,,\label{eq:conditionsstateA1} \\
& ( M_{a, \iota}\otimes \1_{2t}-\1_{2t}\otimes  M_{a, \iota}^*) (\tilde{\mathbb U} \otimes \tilde{\mathbb U}^*) (\tilde{\mathbb W} \otimes \tilde{\mathbb W}^*)\ket{A}=0\,,\label{eq:conditionsstateA2}\\
&
( M_{a, \iota}\otimes \1_{2t}-\1_{2t}\otimes  M_{a, \iota}^*) (\tilde{\mathbb W} \otimes \tilde{\mathbb W}^*) \ket{A} = 0\,,\label{eq:conditionsstateA3}\\
&( M_{a, \iota}\otimes \1_{2t}-\1_{2t}\otimes  M_{a, \iota}^*)\ket{A} = 0\,, \label{eq:conditionsstateA4}
\end{align}
where $\iota\in\{0,1\}$, $a\in\{1,2,\ldots,d^2-1\}$.
\end{lemma} 

\begin{lemma}
\label{lemma7}
For $\tilde{\mathbb U}$ and $\tilde{\mathbb W}$ of the form \eqref{eq:UGE} with $J\neq 0$ and $J'\neq 0$ the conditions \eqref{eq:conditionsstateA2}, \eqref{eq:conditionsstateA3}, and \eqref{eq:conditionsstateA4} are equivalent to 
\be
[A, M_{a,\iota}]=0\,, \quad [A, M_{ab,\iota}]=0\,, \quad a,b\in\{1,2,\ldots,d^2-1\},\quad \iota\in\{0, 1\}\,.
\ee
\end{lemma}

\subsubsection{Proof of Lemma \ref{lemma6}}
\label{sec:prooflemma6}

Considering the expectation value \eqref{eq:EV}, and using that 
\be
\bra{B}\mathbb O^\dagger_{\iota'} \mathbb O^{\phantom{\dag}}_{\iota'}\ket{B} \le \langle B|B\rangle\,,\qquad \bra{B}\mathbb O^{\phantom{\dag}}_{\iota'} \mathbb O^\dagger_{\iota'}\ket{B} \le \langle B|B\rangle\,,
\ee
for any $\ket{B}$, we have 
\bea
1 &=& \braket{A|(\mathbb T_0 \mathbb T)^\dag\mathbb T_0 \mathbb T |A}\nonumber\\
 &=& \| \mathbb O_1^{{\dag}} (\tilde{\mathbb W}_0 \otimes \tilde{\mathbb W}^*_0) \mathbb O_0 (\tilde{\mathbb U} \otimes \tilde{\mathbb U}^*) \mathbb O_1^{{\dag}} (\tilde{\mathbb W} \otimes \tilde{\mathbb W}^*) \mathbb O_0 |A \rangle\|^2   \label{eq:boundxinho} \\
&\le & \| \mathbb O_1^{{\dag}} (\tilde{\mathbb W}_0 \otimes \tilde{\mathbb W}^*_0) \mathbb O_0 (\tilde{\mathbb U} \otimes \tilde{\mathbb U}^*) \mathbb O_1^{{\dag}} (\tilde{\mathbb W} \otimes \tilde{\mathbb W}^*) |A \rangle\|^2  \nonumber \\
&\le & \| \mathbb O_1^{{\dag}} (\tilde{\mathbb W}_0 \otimes \tilde{\mathbb W}^*_0) \mathbb O_0 (\tilde{\mathbb U} \otimes \tilde{\mathbb U}^*) (\tilde{\mathbb W} \otimes \tilde{\mathbb W}^*) |A \rangle\|^2  \nonumber \\
&\le & \| \mathbb O_1^{{\dag}} (\tilde{\mathbb W}_0 \otimes \tilde{\mathbb W}^*_0) (\tilde{\mathbb U} \otimes \tilde{\mathbb U}^*) (\tilde{\mathbb W} \otimes \tilde{\mathbb W}^*) |A \rangle\|^2  \nonumber \\
&\le & \| (\tilde{\mathbb W}_0 \otimes \tilde{\mathbb W}^*_0) (\tilde{\mathbb U} \otimes \tilde{\mathbb U}^*) (\tilde{\mathbb W} \otimes \tilde{\mathbb W}^*) |A \rangle\|^2 = \| | A \rangle\|^2 =1\,.  \nonumber
\eea
This can hold only if all four inequalities in (\ref{eq:boundxinho}) are saturated. Using \eqref{eq:unique} we see that this happens only if the conditions (\ref{eq:conditionsstateA1}\,--\, \ref{eq:conditionsstateA4})  are satisfied.

\subsubsection{Proof of Lemma \ref{lemma7}}

Plugging the forms \eqref{eq:UGE} we bring (\ref{eq:conditionsstateA2}\,--\,\ref{eq:conditionsstateA4}) in the form   
\begin{align}
[M_{a,\iota},A] &\!=\!0, &  [\tilde{\mathbb V}^{\prime \dag} M_{a,\iota} \tilde{\mathbb V}^{\prime},A]&=0, &  [\tilde{\mathbb V}^{\prime \dag} \tilde{\mathbb V}^\dag M_{a,\iota} \tilde{\mathbb V} \tilde{\mathbb V}^{\prime},A]&=0, \label{eq:intermediatecondinho}
\end{align}
where $a\in\{1,2,\ldots,d^2-1\}$, $\iota\in\{0,1\}$. As shown in the proof of Lemma \ref{lemma2} we have that the first two relations \eqref{eq:intermediatecondinho} imply that for $J'\neq 0$:
\be
[A,M_{ab,1}]=0,\qquad  a,b\in\{1,2,\ldots,d^2-1\}.
\label{eq:AdoubleModdinho}
\ee 
Using the above relation and Eq.~\eqref{eq:V'MMMd} we have that the third of \eqref{eq:intermediatecondinho} is equivalent to 
\begin{align}
[\tilde{\mathbb V}^\dag M_{a,\iota} \tilde{\mathbb V},A]=0.
\end{align}
Proceeding as in~(\ref{eq:VMVo}, \ref{eq:AdoubleMeven}) we then find that for $J\neq0$:
\be
[A,M_{ab,0}]=0,\qquad  a,b\in\{1,2,\ldots, d^2-1\}.
\label{eq:AdoubleMeveninho}
\ee

\subsection{Fluctuations of the spectral form factor}
\label{sec:fluctuations}

The approach presented in the current manuscript can also be applied to study the \emph{fluctuations} of the SFF~\eqref{eq:SFF} (this idea has recently been exploited in Ref.~\cite{Fluctuations} in the special case of the self-dual kicked Ising model and in Ref.~\cite{largeq} for the large $d$ asymptotics of Floquet chains with Haar random interactions). Specifically, it can be employed to compute the higher moments of $|{\rm tr}\, \mathbb U_{L}^t|^{2}$ with respect to the 
{\em i.i.d.} on-site disorder distribution 
\be
K_n(t,L) := \mathbb E\left[ |{\rm tr}\, \mathbb U_{L}^t|^{2n}\right] = \mathbb E\left[\left(\sum_{i,j=1}^{\mathcal N} e^{i (\varphi_{i}-\varphi_{j}) t}\right)^{\!\!\!n\;}\right]\!\!,\,\,\, t,L\in\mathbb N,\,\,\, n\geq1.
\label{eq:nSFF}
\ee
Indeed, exploiting the space-time duality described in Sec.~\ref{sec:duality} we can express the above quantities as 
\be
K_n(t,L) = {\rm tr}\,\mathbb T^L_n, 
\ee
with 
\be
\mathbb T_n = (\tilde{\mathbb U}^{\otimes n} \otimes (\tilde{\mathbb U}^*)^{\otimes n}) \mathbb O_{1;n}^\dag (\tilde{\mathbb W}^{\otimes n} \otimes (\tilde{\mathbb W}^*)^{\otimes n}) \mathbb O^{\phantom{\dag}}_{0;n},
\label{eq:SFFTM2n}
\ee
where $\tilde{\mathbb U}$ and $\tilde{\mathbb W}$ are defined in (\ref{eq:Utildee}, \ref{eq:Utildeo}) while we introduced  
\begin{align}
\mathbb O_{\iota';n} &:= \mathbb O_{0\iota';n} \mathbb O_{1\iota';n}=  \mathbb O_{1\iota';n} \mathbb O_{0\iota';n}\,,\\  
\mathbb O_{\iota\iota';n} &:= \int\!{\rm d}^{d^2-1}\boldsymbol{\theta}\,g_{\iota\iota'}(\boldsymbol{\theta})
\exp\left(i\boldsymbol\theta\cdot (\boldsymbol{M}_{\iota;n}^{\phantom{*}}\otimes \1_{2t}-\1_{2t}\otimes \boldsymbol{M}^*_{\iota;n})\right)\,.
\end{align}
and finally $\boldsymbol{M}_{\iota;n} = (M_{1,\iota;n},M_{2,\iota;n},\ldots, M_{d^2-1,\iota;n})$ with 
\be
M_{a, \iota;n} := \sum_{j=0}^{n-1} \1_t^{\otimes j}\otimes\! M_{a, \iota}\!\otimes \1_t^{\otimes (n-1-j)},\quad a\in\{1,\ldots,d^2-1\},\,\, \iota\in\{0,1\},
\ee
where the generalised sublattice magnetisation $M_{a, \iota}$ is defined in Eq.~\eqref{eq:sublatticeM}. Note that, for definiteness, here we considered systems without time-reversal symmetry. 

Using that the matrix $\mathbb T_n$ has the same structure as $\mathbb T$ in \eqref{eq:SFFTM2} we can directly repeat the treatment described in Sec.~\ref{sec:leadingeigs}. In particular, for local gates of the form \eqref{eq:parametrisationgen} we find 
\be
\lim_{L\to\infty} K_n(t,L) = {\rm dim}\, \mathcal M_n',
\ee
where we introduced the set 
\be
\mathcal M_n:= \{ M_{a,\iota; n} \}_{a,\iota} \cup \{ M_{ab,\iota;n} \}_{a,b,\iota}\,,
\label{eq:finalsetn}
\ee
and the coproduct of $n$ double magnetizations (cf. \eqref{eq:doubleM}) 
\be
M_{a b,\iota; n}:= \!\sum_{j=1}^n\! \1_t^{\otimes (j-1)}\otimes\! M_{ab, \iota}\!\otimes \1_t^{\otimes (n-j)},\,\,  a,b\in\{1,\ldots,d^2-1\},\,\iota\in\{0,1\}\,.
\label{eq:doubleMn}
\ee
All elements of \eqref{eq:finalsetn} are invariant under permutations of the $n$ copies of $\mathcal H_t$ in $\mathcal H_t^{\otimes n}$ and under 2-site translations within each copy, i.e., they commute with  
\begin{align}
A_{p; \tau_1, \ldots \tau_n} &= \Gamma(p) (\Pi_{2t}^{2 \tau_1} \otimes\cdots\otimes \Pi_{2t}^{2 \tau_n}), \quad \tau_1,\ldots, \tau_n\in \{0,\ldots,t-1\}\,, 
\end{align}
where $\Gamma(\cdot)$ is a representation of $S_n$, the symmetric group of $n$ letters, on $\mathcal H_{t n}\equiv \mathcal H_t^{\otimes n}$ and $p\in S_n$. Specifically,
\be
\Gamma(p) \ket{A_1}\otimes\ket{A_2}\otimes\cdots\otimes \ket{A_n} = \ket{A_{p(1)}}\otimes\ket{A_{p(2)}}\otimes\cdots\otimes \ket{A_{p(n)}}.
\ee
This means that 
\be
\mathcal A_n = {\rm span}\{ A_{p; \tau_1,\tau_2 \ldots \tau_n};\,\, \tau_1,\tau_2 \ldots, \tau_n\in \{0,\ldots,t-1\},\, p\in S_n\}
\ee
is a vector subspace of $\mathcal M_n'$ and hence 
\be
\lim_{L\to\infty} K_n(t,L)={\rm dim}\, \mathcal M_n'\geq n!\cdot t^n =   \lim_{\mathcal N \to\infty} \int |{\rm tr}\,\mathbb{U}^t|^{2n} {\rm d}\mu_{\rm CUE}(\mathbb{U}),
\label{eq:fluctuations}
\ee
where ${\rm d}\mu_{\rm CUE}(\mathbb{U})$ is the CUE measure and $\mathcal N$ is the dimension of the matrix $\mathbb{U}$. 

In analogy with what happens for $n=1$ (c.f. Theorem~\ref{theorem1}) we expect $\mathcal A_n$ and $\mathcal M_n'$ to actually coincide, leading to an equality sign in \eqref{eq:fluctuations}. However, we leave the formal proof of this statement to future work. Similar conclusions (with CUE replaced by COE) hold in the time-reversal invariant case.

\subsection{Singular on-site disorder distributions}
\label{sec:nonisotropic}

As discussed in Sec.~\ref{sec:SFF} we expect our results to be stable under modifications of the averaging procedure as long as such modifications do not introduce spatial correlations. In our setting this can be verified explicitly by considering \emph{singular} disorder distributions of local gates $u_x,w_x$, Eq.~\eqref{eq:urx}, supported on lower-dimensional submanifolds of ${\rm SU}(d)$ that include the identity. For instance, one can imagine
 having some of the components of $\boldsymbol{\theta}_{\iota,x}$ in \eqref{eq:urx} (or $\boldsymbol{\theta}_{x}$ in~\eqref{eq:urxT}) set to zero for all $\iota$ and $x$. Physically, this choice describes a weaker external noise where, for example, the random magnetic fields are imposed only along certain specific directions rather than isotropically.   

In this case we expect that away from certain ``resonances", namely for almost all 2-site dual-unitary gates $U, W$, the treatment described in the previous sections is still applicable. In particular we anticipate that the SFF will still be characterised by the commutants of \eqref{eq:commutant} or \eqref{eq:commutantT} depending on whether or not the problem is time-reversal symmetric. 

Let us illustrate the main steps that can be used to prove this idea. We consider for simplicity the case of qubits ($d=2$) and absence of time-reversal symmetry. Denoting by $\mathcal I$ the subset of indices $\mathcal I\subset\{1,2,3\}$ such that $\{\theta_{a, \iota, x}\}_{a\in \mathcal I}$ are \emph{not} set to zero, and repeating the steps of Sec.~\ref{sec:duality} and \ref{sec:leadingeigs}, one readily finds the following analogue of the conditions \eqref{eq:CondSU2} 
\be
\begin{split}
&\tilde{\mathbb U} \tilde{\mathbb W} A \tilde{\mathbb W}^\dag \tilde{\mathbb U}^\dag =e^{i \phi} A\,,\qquad [M_{a, \iota},A] = 0\,,\\
& [\tilde{\mathbb W}^\dag M_{a, \iota} \tilde{\mathbb W},A]=0\,,\qquad\qquad\qquad\qquad\qquad \iota\in\{0,1\}\,,\,\, a\in\mathcal I\,.  
\end{split}
\label{eq:CondSU1}
\ee
At this point we note that if $\mathcal I $ has at least two elements these conditions are equivalent to  \eqref{eq:CondSU2}. This follows by observing that if $A$ commutes with $M_{a, \iota}$ and $M_{b, \iota}$ it also commutes with their commutator
\be
[M_{a, \iota},M_{b, \iota}] = i \sum_{c=1}^3\epsilon_{abc} M_{c, \iota}\,. 
\ee
Here we used that $\{M_{a, \iota}\}$ generate the $\mathfrak{su}(2)$ algebra. A completely analogous reasoning applies for the set of equivalent matrices
$\{\tilde{\mathbb W}^\dag M_{a, \iota} \tilde{\mathbb W}\}$. 

The only possible non-trivial choice is then to take the set $\mathcal I$ composed by single element, which, without loss of generality, can be set to $3$. This corresponds to $u_x,w_x$ being restricted to some ${\rm U}(1)$ subgroup of ${\rm SU}(2)$. To show that the our treatment applies also in this case we need to prove the analogue of Lemma~\ref{lemma2}. Namely, we need to show that if $A$ fulfils \eqref{eq:CondSU1} then it commutes with $\{M_{a,\iota}\}_{a=1,2,3; \iota=0,1}$, $\{M_{ab,\iota}\}_{a,b=1,2,3; \iota=0,1}$. 
This can be done by considering the following set of $15$ operators
\be
 \begin{split}
\mathcal S_1 =&\{ \{M_{3, \iota}\}_{\iota=0,1},  \{[N_{\iota}, M_{3, \iota'}]\}_{\iota,\iota'=0,1},
 \{[[N_\iota, M_{3, \iota'}], M_{3, \iota'}]\}_{\iota,\iota'=0,1},  \\
&\{N_{\iota}\}_{\iota=0,1}, [[N_0, M_{3, 0}], N_0],  [[N_0, M_{3,1}], N_0], [[N_1, M_{3,0}], N_1]
 \} \, ,
 \end{split}
\label{eq:U1ops}
\ee
where we introduced the short-hand notation 
\be
N_\iota:=\tilde{\mathbb W}^\dag M_{3, \iota} \tilde{\mathbb W}^{\phantom{\dag}}\!\!, 
\quad \iota=0,1.
\ee
Since the operators \eqref{eq:U1ops} are constructed by taking commutators of $M_{3, \iota}$, and $\tilde{\mathbb W}^\dag M_{3, \iota} \tilde{\mathbb W}$, 
they commute with $A$. Moreover, they can be written as linear combinations of $\{M_{a,\iota}\}_{a=1,2,3; \iota=0,1}$ and $\{M_{ab,1}\}_{a,b=1,2,3}$. This means that if we can prove that the operators in $\mathcal S_1$ are \emph{linearly independent} we immediately have  
\be
[A, M_{a, \iota}]=0,\qquad [A, M_{ab, 1}]=0\,,\qquad  a,b\in\{1,2,3\},\qquad \iota\in\{0,1\}\,.
\ee
We could not prove explicitly the linear independence of the set \eqref{eq:U1ops} but we verified numerically that it holds almost always (away from special, measure-zero set of $U$, $W$ --- the so-called ``resonances'').  

An analogous reasoning considering the set of $15$ operators 
\be
 \begin{split}
\mathcal S_2 =&\{ \{[[\tilde N_\iota, M_{3, \iota'}], M_{3, \iota'}]\}_{\iota,\iota'=0,1}, \{\tilde N_{\iota}\}_{\iota=0,1}, [[\tilde N_0, M_{3, 0}], \tilde N_0],\\
& [[\tilde N_0, M_{3,1}], \tilde N_0], \{[\tilde N_{\iota}, M_{3, \iota'}]\}_{\iota,\iota'=0,1}, [[\tilde N_1, M_{3,0}], \tilde N_1]
 \} \, ,
 \end{split}
\label{eq:U1ops2}
\ee
with 
\be
\tilde{N}_\iota:=\tilde{\mathbb U} \tilde{\mathbb W}  \tilde{\mathbb W}^\dag M_{3, \iota} \tilde{\mathbb W} \tilde{\mathbb W}^\dag \tilde{\mathbb U}^\dag=\tilde{\mathbb U} M_{3, \iota} {\tilde{\mathbb U}}^\dag\,,  
\ee
leads to 
\be
[A, M_{a, \iota}]=0,\qquad [A, M_{ab, 0}]=0\,,\qquad a,b\in\{1,2,3\},\qquad \iota\in\{0,1\}\,.
\ee
This means that
\be
\lim_{L\to\infty} K(t,L) = \dim \mathcal M' = t, 
\ee
where in the second step we applied Theorem~\ref{theorem1}.

\begin{acknowledgements}
This work has been supported by the EU Horizon 2020 program through the European Research Council (ERC) Advanced Grant OMNES No. 694544, by the Slovenian Research Agency (ARRS) under the Programme P1-0402, and by the Royal Society through the University Research Fellowship No. 201101 (BB). BB acknowledges useful discussions with Andrea De Luca.
\end{acknowledgements}

%
%



\end{document}